\def\be{\begin{equation}}
	\def\ee{\end{equation}}
\def\ba{\begin{array}}
	\def\ea{\end{array}}

\def\mathbi#1{\text{\em #1}}

\documentclass[prl,showpacs,twocolumn,amsmath]{revtex4}
\usepackage{amsmath}
\usepackage{amsfonts}
\usepackage{mathrsfs}
\usepackage{amssymb}
\usepackage{pifont}
\usepackage{epsfig,subfigure,dsfont,amsthm,amsbsy,mathrsfs,amscd}
\usepackage{epstopdf}
\usepackage{bbm}
\usepackage{color}
\usepackage{autobreak}
\def\qed{\leavevmode\unskip\penalty9999 \hbox{}\nobreak\hfil
	\quad\hbox{\leavevmode  \hbox to.77778em{%
			\hfil\vrule   \vbox to.675em%
			{\hrule width.6em\vfil\hrule}\vrule\hfil}}
	\par\vskip3pt}
\usepackage{leftidx}
\newtheorem{theorem}{Theorem}

\input amssym.def
\begin{document}
\title{\large\bf Trade-off relations and enhancement protocol of quantum battery capacities in multipartite systems}
	
\author{Yiding Wang$^{1}$, Xiaofen Huang$^{1}$, Shao-Ming Fei$^{2}$ and Tinggui Zhang$^{1,\dag}$}
	\affiliation{$1$ School of Mathematics and Statistics, Hainan Normal University, Haikou, 571158, China \\$2$ School of Mathematical Sciences, Capital Normal University, Beijing 100048, China\\
		\\  $^{\dag}$ Correspondence to tinggui333@163.com }

	\bigskip
	
\begin{abstract}
First, we investigate the trade-off relations of quantum battery capacities in two-qubit system. We find that the sum of subsystem battery capacity is governed by the total system capacity, with this trade-off relation persisting for a class of Hamiltonians, including Ising, XX, XXZ and XXX models. Then building on this relation, we define residual battery capacity for general quantum states and establish coherent/incoherent components of subsystem battery capacity. Furthermore, we introduce the protocol to guide the selection of appropriate incoherent unitary operations for enhancing subsystem battery capacity in specific scenarios, along with a sufficient condition for achieving subsystem capacity gain through unitary operation. Numerical examples validate the feasibility of the incoherent operation protocol. Additionally, for the three-qubit system, we also established a set of theories and results parallel to those for two-qubit case. Finally, we determine the minimum time required to enhance subsystem battery capacity via a single incoherent operation in our protocol. Our findings contribute to the development of quantum battery theory and quantum energy storage systems.

\end{abstract}
	

	\bigskip
	
\begin{abstract}
First, we investigate the trade-off relations of quantum battery capacities in two-qubit system. We find that the sum of subsystem battery capacity is governed by the total system capacity, with this trade-off relation persisting for a class of Hamiltonians, including Ising, XX, XXZ and XXX models. Then building on this relation, we define residual battery capacity for general quantum states and establish coherent/incoherent components of subsystem battery capacity. Furthermore, we introduce the protocol to guide the selection of appropriate incoherent unitary operations for enhancing subsystem battery capacity in specific scenarios, along with a sufficient condition for achieving subsystem capacity gain through unitary operation. Numerical examples validate the feasibility of the incoherent operation protocol. Additionally, for the three-qubit system, we also established a set of theories and results parallel to those for two-qubit case. Finally, we determine the minimum time required to enhance subsystem battery capacity via a single incoherent operation in our protocol. Our findings contribute to the development of quantum battery theory and quantum energy storage systems.
\end{abstract}
	
 \maketitle
	
\section{I. Introduction} 
In the last decades, the energetics of quantum systems has been extensively explored within the emerging fields of quantum information and quantum thermodynamics \cite{ihrm,kmjp,pacj,imsv}. In this context, the concept of quantum battery was first introduced by Alicki and Fannes \cite{ramf}, in which they investigated how quantum resources such as entanglement and coherence could enable efficient work extraction from quantum systems. Following this ground-breaking work, significant efforts have been devoted to investigating quantum batteries, with various theoretical models being explored for identifying optimal protocols of charging and discharging \cite{kvhm,gmfp,nlcb,dfmc,wljc,mlsl,yyxq,fyja,khhg,fmvc,strss,agcs,imad,snem,rcrn,zgxc,szzz,yyxs,lsfh}. Among these theoretical models are many-body quantum batteries, for which the Hamiltonians feature interactions among the subsystems. This type of models is referred to as spin-chain quantum batteries. The first spin-chain quantum battery model was introduced by Le et al. \cite{tplj}, in which it was shown that the spin-spin interactions yield an advantage in charging power over the non-interacting case, and such advantage grows super-extensively when the interactions are long ranged. Since then numerous studies have emerged focusing on spin-chain quantum battery models \cite{yyxq,imad,sgtc,sgas,fbkv,fqdh,xlzx,aasak}. The authors in \cite{yyxq} investigated the problem of charging a dissipative one-dimensional XXX spin-chain quantum battery by using local magnetic fields in the presence of spin decay. Inspired by the variational quantum eigensolver (VQE) algorithm Medina et al. proposed an approach to optimize the extractable energy \cite{imad}. In \cite{xlzx} the authors considered the characteristics of quantum batteries for Heisenberg spin chain models in the absence and presence of Dzyaloshinskii-Moriya (DM) interaction and showed that the first-order coherence is a crucial quantum resource during charging. Recent progress in quantum battery research has been highlighted in a comprehensive review \cite{fcsg}.

One of central quantity in theory of quantum battery is the ergotropy \cite{ramf,aear,assd}, defined as the maximum amount of energy that can be extracted from a quantum system through unitary operations \cite{gffc,fhks,gf,kkky,dthf,rcdf,rpas,spjb,rcrn}. Notably, Yang et al. proposed a new definition of quantum battery capacity \cite{yyas}, which is verified experimentally based on optical platforms \cite{xyyh}. This quantum battery capacity has been further investigated with respect to different scenarios \cite{aasak,tgzh,ykwl,wyd}. However,
it remains as highly valuable problems to address the current knowledge gap regarding the trade-off relations of quantum battery capacity, and establish systematic protocols based on these trade-off relations to enhance the subsystem's battery capacity without weakening the whole system's capacity, as well as to determine the minimum time required to enhance the subsystem's battery capacity via a single incoherent operation. We provide the trade-off relation of quantum battery capacity in two-qubit system and the sufficient condition for achieving the subsystem's capacity enhancement with incoherent operation protocol, and present the minimum time required to enhance the subsystem's battery capacity via incoherent operations. 

The rest of this paper is organized as follows. In the section II, we provide main results of this paper, including the trade-off relations of quantum battery capacity in two- and three-qubit systems [Theorem 1 and 3], the global unitary operation protocol for two-qubit system, and the sufficient conditions for achieving subsystem capacity enhancement through our global unitary operation protocol [Theorem 2 and 4]. In section III, we present the minimum time required to enhance subsystem battery capacity via incoherent operations in our protocol [Theorem 5]. In addition, the proofs of these theorems, the incoherent unitary operation protocols in three-qubit quantum systems, and the calculation details of numerical examples are included in the Appendix. We summarize and discuss our conclusions in the last section.

\section{II. Trade-off relations of quantum battery capacity}
 The quantum battery capacity of a $d$-dimensional battery state $\rho$ is defined by \cite{yyas}
\begin{equation}\label{e1}
	\begin{split}
		\mathcal{C}(\rho;H)&=\sum_{i=0}^{d-1}\epsilon_i(\lambda_i-\lambda_{d-1-i})
	\end{split}
\end{equation}
with respect to a given Hamiltonian $H=\sum_{i=0}^{d-1}\epsilon_i|\varepsilon_i\rangle\langle\varepsilon_i|$,
where $\{\lambda_i\}$ ($\{\epsilon_i\}$) represent the energy levels of $\rho$ ($H$), arranged in ascending order, $\lambda_0\leqslant\lambda_1\leqslant...\leqslant\lambda_{d-1}$ and $\epsilon_0\leqslant\epsilon_1\leqslant...\leqslant\epsilon_{d-1}$. The quantum battery capacity $\mathcal{C}$ is a Schur-convex functional of $\rho$, i.e., if $\rho$ is majorized by $\varrho$ $(\rho\prec\varrho)$, then $\mathcal{C}(\rho;H)\leq\mathcal{C}(\varrho;H)$. 

We consider a two-qubit quantum battery consisting two coupled two-level systems $A-B$. Without loss of generality, the Hamiltonian of the whole quantum battery system is given by \cite{imad,tplj,xlzx}
\begin{equation}\label{e2}
\begin{split}
H&=H_0+H_\text{int}.
\end{split}
\end{equation}
The first term $H_0$ characterizes the external magnetic field. Here, we can consider two types of $H_0$:
\begin{equation*}
	\begin{split}
		H_0&=E(\sigma_x\otimes I_2+\sigma_y\otimes I_2+I_2\otimes\sigma_x+I_2\otimes\sigma_y),\\
		H_0&=E(\sigma_z\otimes I_2+I_2\otimes\sigma_z),
	\end{split}
\end{equation*}
representing the transverse and longitudinal external magnetic fields, respectively, where $E$ denotes the magnetic field strength and $\sigma_i$ $(i=x,y,z)$ are the standard Pauli matrices. The corresponding subsystems' Hamiltonian are $H_A=H_B=E(\sigma_x+\sigma_y)$ and $H_A=H_B=E\sigma_z$, respectively. The last term in Hamiltonian, $H_\text{int}$, defines the interactions among the spins.

From Eq.\,(\ref{e1}), it can be seen that the battery capacity trade-off relation is intimately connected to the problem of the compatibility of the spectrum of local states with the spectrum of the global state, which is so-called quantum marginal problem (QMP) \cite{sbr,rpas,mcgm,aak}. For two qubits system, QMP is about the relations between the spectrum of the mixed state $\rho_{AB}$ of two component system $A-B$ and that of reduced states $\rho_A$ and $\rho_B$. The reduced states $\rho_A$ and $\rho_B$ with spectra $\{\lambda_0^A\leq\lambda_1^A\}$ and $\{\lambda_0^B\leq\lambda_1^B\}$ can be the marginals of a global state $\rho_{AB}$ with the spectrum $\{\lambda_0\leq\lambda_1\leq\lambda_2\leq\lambda_3\}$ only if \cite{sbr,rpas}
\begin{align}
	\lambda_0^A&\geq\lambda_0+\lambda_1,\label{e3}\\
	\lambda_0^B&\geq\lambda_0+\lambda_1,\label{e4}\\
	\lambda_0^A+\lambda_0^B&\geq2\lambda_0+\lambda_1+\lambda_2,\label{e5}\\
	|\lambda_0^A-\lambda_0^B|&\leq\min\{\lambda_3-\lambda_1,\lambda_2-\lambda_0\}.\label{e6}
\end{align}

Using QMP, we can obtain the following result.
\begin{theorem}
For any two-qubit quantum state $\rho$, the following trade-off relation holds:
\begin{equation}\label{e7}
\mathcal{C}(\rho_A;H_A)+\mathcal{C}(\rho_B;H_B)\leq\mathcal{C}(\rho;H),
\end{equation}
for any Hamiltonian $H=H_0+H_{int}$ such that $\mathcal{C}(\rho;H)\geq\mathcal{C}(\rho;H_0)$,
where $\rho_A$ and $\rho_B$ are reduced density matrices of $\rho$.
\end{theorem}

See Appendix A for the proof of Theorem 1. We illustrate Theorem 1 by considering the following Hamiltonian \cite{imad,tplj,xlzx}, $H=H_0+\alpha J(\sigma_x\otimes\sigma_x+\sigma_y\otimes\sigma_y)+\beta J\sigma_z\otimes\sigma_z$, where $J$ stands for the interaction strength, $\alpha\,(|\alpha|\leq1)$ and $\beta\,(|\beta|\leq1)$ represent the anisotropy. The trade-off relation of quantum battery capacity given by (\ref{e7}) holds universally for the Ising ($J>0$, $\alpha=0$, $\beta=1$), XXZ ($J>0$, $\alpha\neq0$, $\beta=1$), XX ($J>0$, $\alpha\neq0$, $\beta=0$), and XXX models ($J>0$, $\alpha=\beta=1$), see Appendix B.

Theorem 1 implies that for a generic two-qubit state, the sum of the subsystem battery capacity cannot exceed the total system capacity. So we can define the residual battery capacity ($\triangle(\rho,H)$) for general two-qubit state $\rho$,
\begin{equation}\label{e8}
\triangle(\rho,H)=\mathcal{C}(\rho;H)-[\mathcal{C}(\rho_A;H_A)+\mathcal{C}(\rho_B;H_B)].
\end{equation}
For simplicity, we denote $Sub(\rho)$ the subsystems' battery capacity $\mathcal{C}(\rho_A;H_A)+\mathcal{C}(\rho_B;H_B)$. 

We naturally desire that the subsystem holds a significant share of the total battery capacity in a quantum state. However, many states, including the Bell states, possess high total battery capacity, yet their subsystems have zero capacity. Therefore, it is of practical significance to manipulate the quantum battery capacity distributions, for instance, such that the subsystems account for a high proportion the in quantum battery capacity distribution. In this context, global unitary operation may be a suitable choice. Generally, the local evolutions of the subsystems $A$ and $B$ induced by a global unitary evolution $U_{AB}$ are not unitary \cite{fgsl,ja,javs}. Because of this, the global unitary operation changes the spectral structure of the reduced states while keeping the entire system battery capacity unchanged, which may improve the battery capacity distribution relationship. Let $U$ be a global unitary operator, and $\Tilde{\rho}_A$ and $\Tilde{\rho}_B$ the reduced states of $U\rho\,U^\dagger$. We say that this process increases the subsystems' capacity $Sub(\rho)$ if $$\mathcal{C}(\Tilde{\rho}_A;H_A)+\mathcal{C}(\Tilde{\rho}_B;H_B)
-\mathcal{C}(\rho_A;H_A)-\mathcal{C}(\rho_B;H_B)>0.$$

We investigate the increase of $Sub(\rho)$ from the contributions of quantum coherent and incoherent parts related to the subsystems' capacities. The incoherent part $Sub_{ic}(\rho)$ of the subsystems' capacities is obtained by applying a dephasing map that completely erases the coherence of $\rho$,
\begin{equation}\label{e9}
Sub_{ic}(\rho)=\mathcal{C}(\tau_A;H_A)+\mathcal{C}(\tau_B;H_B),
\end{equation} 
where $\tau$ is the dephased state of $\rho$, $\tau_A$ and $\tau_B$ are the reduced states of $\tau$. The coherent contribution to the subsystems' capacities is then given by
\begin{equation}\label{e10}
Sub_c(\rho)=Sub(\rho)-Sub_{ic}(\rho).
\end{equation}


We show next that the incoherent part of the subsystems' battery capacities can be enhanced via global unitary operations, see proof in Appendix C.

\begin{theorem}
For any two-qubit quantum state, global unitary operations can enhance the value of $Sub_{ic}(\rho)$. In particular, the maximum eigenvalues of $\rho_A$, $\rho_B$, $\Tilde{\tau}_A$ and $\Tilde{\tau}_B$ are denoted as $\lambda_1^A$, $\lambda_1^B$, $\xi_1^A$ and $\xi_1^B$, respectively. If the unitary operation $U$ makes 
\begin{equation}\label{e11}
	\xi_1^A+\xi_1^B>\lambda_1^A+\lambda_1^B,
\end{equation}
then this process achieves the subsystem capacity gain.
\end{theorem}

Notably, the operations involved in Theorem 2 are not only unitary but also belong to the class of incoherent operations \cite{tbmc}, which are free operations within the resource theory of quantum coherence. Additionally, our framework can also be applied to generic quantum battery models by designating one qubit as the charger and the rest as the battery. In this configuration, the enhanced capacity of the subsystems directly increases the capacity of the battery. The process of using global unitary operations to improve the battery capacity distribution is shown in Figure 1.
\begin{figure}[htbp]
	\centering
	\includegraphics[width=0.5\textwidth]{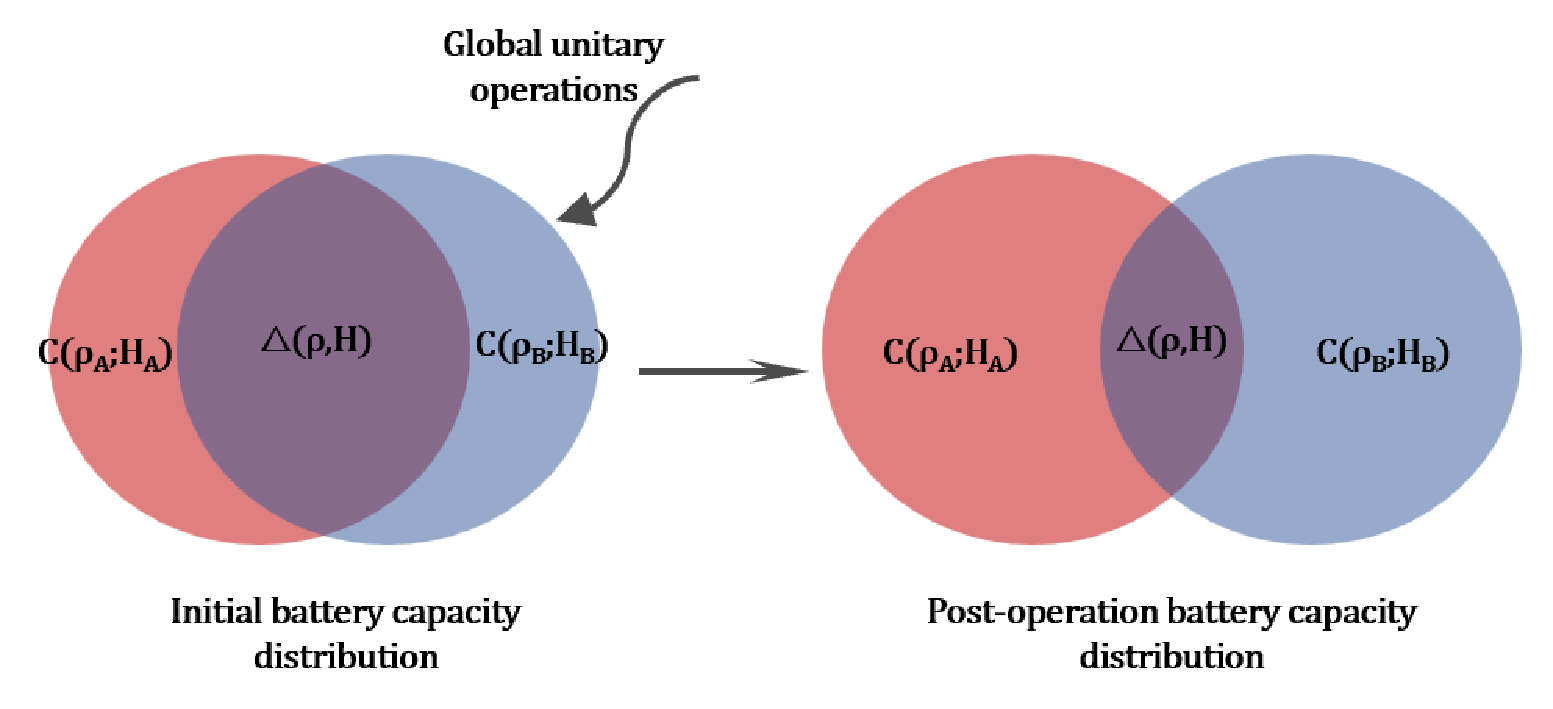}
	\vspace{-1em} \caption{Our approach involves applying global unitary operations to reduce the value of $\triangle(\rho,H)$, thereby increasing the value of $Sub(\rho)$, since $\mathcal{C}(\rho;H)$ does not change under the global unitary operation.} \label{Fig.1}
\end{figure}

Increasing the value of $Sub_{ic}(\rho)$ can enhance the subsystems' capacities for $X$-state \cite{wyd}. However, for general state,  the subsystem capacity is not solely dependent on $Sub_{ic}(\rho)$, but also the coherent components within the reduced density matrix. This implies that the increase of $Sub_{ic}(\rho)$ cannot guarantee the subsystem capacity gain. In fact, this is a state-dependent problem. A natural idea is to introduce an incoherent unitary operation protocol, illustrating the way to select appropriate unitary operations to achieve subsystem capacity gain.

Given a two-qubit state $\rho=(\rho_{ij})_{4\times4}$, the eigenvalues of reduced states $\rho_A$ and $\rho_B$ are
\begin{equation*}
	\begin{split}
		\lambda_0^A&=\frac{1}{2}-\frac{\sqrt{4|\rho_{13}+\rho_{24}|^2+(\rho_{11}+\rho_{22}-\rho_{33}-\rho_{44})^2}}{2},\\
		\lambda_1^A&=\frac{1}{2}+\frac{\sqrt{4|\rho_{13}+\rho_{24}|^2+(\rho_{11}+\rho_{22}-\rho_{33}-\rho_{44})^2}}{2},\\
		\lambda_0^B&=\frac{1}{2}-\frac{\sqrt{4|\rho_{12}+\rho_{34}|^2+(\rho_{11}+\rho_{33}-\rho_{22}-\rho_{44})^2}}{2},\\
		\lambda_1^B&=\frac{1}{2}+\frac{\sqrt{4|\rho_{12}+\rho_{34}|^2+(\rho_{11}+\rho_{33}-\rho_{22}-\rho_{44})^2}}{2}.
	\end{split}
\end{equation*}
Since our protocol is Hamiltonian-independent, we set the energy levels of $H_A$ and $H_B$ as $\varepsilon_0^X\leq\varepsilon_1^X\,(X=A,B)$. Then one has
\begin{small}
\begin{equation*}
	\begin{split}
		\mathcal{C}(\rho_A;H_A)&=(\varepsilon_1^A-\varepsilon_0^A)\sqrt{4|\rho_{13}+\rho_{24}|^2+(\rho_{11}+\rho_{22}-\rho_{33}-\rho_{44})^2},\\
		\mathcal{C}(\rho_B;H_B)&=(\varepsilon_1^B-\varepsilon_0^B)\sqrt{4|\rho_{12}+\rho_{34}|^2+(\rho_{11}+\rho_{33}-\rho_{22}-\rho_{44})^2}.
	\end{split}
\end{equation*}
\end{small}
According to the expressions of $\mathcal{C}(\rho_A;H_A)$ and $\mathcal{C}(\rho_B;H_B)$, the battery capacities of the subsystems are given by the coherent and incoherent parts. Denote $2|\rho_{13}+\rho_{24}|$ and $2|\rho_{12}+\rho_{34}|$ as $\text{C}_A$ and $\text{C}_B$ with respect the coherent part, and $(\rho_{11}+\rho_{22}-\rho_{33}-\rho_{44})$ and $(\rho_{11}+\rho_{33}-\rho_{22}-\rho_{44})$ as $\text{IC}_A$ and $\text{IC}_B$ associated with the incoherent part. The battery capacity of the subsystem $A$ can be rewritten as
\begin{equation}
	\begin{split}\mathcal{C}(\rho_A;H_A)&=(\varepsilon_1^A-\varepsilon_0^A)
		\sqrt{\text{C}_A^2+\text{IC}_A^2},\\
		\mathcal{C}(\rho_B;H_B)&=(\varepsilon_1^B-\varepsilon_0^B)\sqrt{\text{C}_B^2+\text{IC}_B^2}.
	\end{split}
\end{equation}

The unitary matrices involved in the protocol are actually a class of incoherent operations. For example, $U_{ij}$ is the matrix obtained by exchanging the $i$th row and the $j$th row of the identity matrix. We illustrate these unitary matrices in the battery capacity distribution one by one.

(1)\,$U_{14}$ and $U_{23}$. They exchange the battery capacity of subsystems $A$ and $B$.
\begin{proof}
	We only verify $U_{14}$.
	$$\Tilde{\rho}=U_{14}\rho\,U_{14}^\dagger=\left(
	\begin{array}{cccc}
		\rho_{44} & \rho_{24}^* & \rho_{34}^* & \rho_{14}^*\\
		\rho_{24} & \rho_{22} & \rho_{23} & \rho_{12}^*\\
		\rho_{34} & \rho_{23}^* & \rho_{33} & \rho_{13}^*\\
		\rho_{14} & \rho_{12} & \rho_{13} & \rho_{11}\\
	\end{array}
	\right ).$$
	So the maximum eigenvalues of $\Tilde{\rho}_A$ and $\Tilde{\rho}_B$ are
	\begin{equation*}
		\begin{split}
			\Tilde{\lambda}_1^A&=\frac{1}{2}+\frac{\sqrt{4|\rho_{12}+\rho_{34}|^2+(\rho_{44}+\rho_{22}-\rho_{33}-\rho_{11})^2}}{2},\\
			\Tilde{\lambda}_1^B&=\frac{1}{2}+\frac{\sqrt{4|\rho_{13}+\rho_{24}|^2+(\rho_{44}+\rho_{33}-\rho_{22}-\rho_{11})^2}}{2}.
		\end{split}
	\end{equation*}
	It is easy to see that $\Tilde{\lambda}_1^A=\lambda_1^B$ and $\Tilde{\lambda}_1^B=\lambda_1^A$. Hence, the battery capacities of $A$ and $B$ are exchanged.
\end{proof}

(2)\,$U_{12}$ and $U_{34}$. The keep $\text{IC}_A$ unchanged and exchange the values of $\text{C}_A$ and $\text{C}^*$.
\begin{proof}
	We only verify $U_{12}$ (the case of $U_{34}$ is similarly verified). We have
	$$\Tilde{\rho}=U_{12}\rho\,U_{12}^\dagger=\left(
	\begin{array}{cccc}
		\rho_{22} & \rho_{12}^* & \rho_{23} & \rho_{24}\\
		\rho_{12} & \rho_{11} & \rho_{13} & \rho_{14}\\
		\rho_{23}^* & \rho_{13}^* & \rho_{33} & \rho_{34}\\
		\rho_{24}^* & \rho_{14}^* & \rho_{34}^* & \rho_{44}\\
	\end{array}
	\right ).$$
	Then the reduced state of the subsystem A is
	$$\Tilde{\rho}_A=\left(
	\begin{array}{cc}
		\rho_{11}+\rho_{22} & \rho_{14}+\rho_{23} \\
		\rho_{14}^*+\rho_{23}^* & \rho_{33}+\rho_{44} \\
	\end{array}	
	\right).$$
	From (\ref{e1}) the subsystem's capacity is
	\begin{small}
	\begin{equation*}
		\begin{split}
			\mathcal{C}(\Tilde{\rho}_A;H_A)&=(\varepsilon_1^A-\varepsilon_0^A)\sqrt{4|\rho_{14}
				+\rho_{23}|^2+(\rho_{11}+\rho_{22}-\rho_{33}-\rho_{44})^2}\\
			&=(\varepsilon_1^A-\varepsilon_0^A)\sqrt{\text{C}^{*2}+\text{IC}^2}.
		\end{split}
	\end{equation*}
	\end{small}
	Note that this process may change the values of $\text{IC}_B$ and $\text{C}_B$.
\end{proof}

(3)\,$U_{13}$ and $U_{24}$. They keep $\text{IC}_B$ unchanged and exchange the values of $\text{C}_B$ and $\text{C}^*$. The proof is similar to that of $U_{12}$.

According to the expressions of $\mathcal{C}(\rho_A;H_A)$ and $\mathcal{C}(\rho_B;H_B)$, we find that the subsystem battery capacity is affected by the incoherent part, that is, the ordering of elements on the diagonal of the total system density matrix. Assume that the diagonal elements are $\mu_1\leq\mu_2\leq\mu_3\leq\mu_4$ in ascending order. The maximum value that $\text{IC}_A$ and $\text{IC}_B$ can reach is $\mu_4+\mu_3-\mu_2-\mu_1$. In general, $\text{IC}_A$ and $\text{IC}_B$ cannot be maximized simultaneously. However, it is possible to make one of them the largest and another the second largest. The optimal ordering of $\text{IC}_A$ ($\text{IC}_B$) is the one with the largest $\text{IC}_A$ ($\text{IC}_B$) value and the second largest $\text{IC}_B$ ($\text{IC}_A$) value. We list the optimal ordering for $\text{IC}_A$ and $\text{IC}_B$ respectively below.

{\bf Optimal ordering for $\text{IC}_A$:}
\begin{equation}
	\begin{split}
		&\rho_{11}\geq\rho_{22}\geq\rho_{33}\geq\rho_{44},\,\,\,\,\rho_{22}\geq\rho_{11}\geq\rho_{44}\geq\rho_{33},\\
		&\rho_{33}\geq\rho_{44}\geq\rho_{11}\geq\rho_{22},\,\,\,\,\rho_{44}\geq\rho_{33}\geq\rho_{22}\geq\rho_{11}.
	\end{split}
\end{equation}

{\bf Optimal ordering for $\text{IC}_B$:}
\begin{equation}
	\begin{split}
		&\rho_{11}\geq\rho_{33}\geq\rho_{22}\geq\rho_{44},\,\,\,\,
		\rho_{22}\geq\rho_{44}\geq\rho_{11}\geq\rho_{33},\\
		&\rho_{33}\geq\rho_{11}\geq\rho_{44}\geq\rho_{22},\,\,\,\,
		\rho_{44}\geq\rho_{22}\geq\rho_{33}\geq\rho_{11}.
	\end{split}
\end{equation}

It is easy to verify that for the ordering in (13), we have
\begin{equation*}
	\begin{split}
		\text{IC}_A&=\mu_4+\mu_3-\mu_2-\mu_1,\,\,\,\,
		\text{IC}_B=\mu_4+\mu_2-\mu_3-\mu_1,
	\end{split}
\end{equation*}
and for (14),
\begin{equation*}
	\begin{split}
		\text{IC}_B&=\mu_4+\mu_3-\mu_2-\mu_1,\,\,\,\,
		\text{IC}_A=\mu_4+\mu_2-\mu_3-\mu_1.\\
	\end{split}
\end{equation*}

We now introduce our incoherent operation protocol. Given a two-qubit battery state $\rho$, this technical protocol is divided into four steps.

{\bf Step 1.} Calculate the residual battery capacity $\triangle(\rho,H)$. If $\triangle(\rho,H)=0$, the protocol ends. Otherwise, go to the next step.

{\bf Step 2.} Calculate the value of $\mathcal{C}(\rho_A;H_A)+\mathcal{C}(\rho_B;H_B)$ and record it as $\text{c}_1$. Determine whether the diagonal ordering belongs to (13) or (14). If the ordering belongs to (13) or (14), go to next step. Otherwise, we use the unitary matrices $U_{ij}$ $(1\leq i<j\leq4)$ to adjust the diagonal ordering to the optimal one belonging to (13) or (14). Calculate the sum of the subsystems' capacities and record it as $\text{c}_2$, and go to the next step.

{\bf Step 3.} At this time, the diagonal ordering of the density matrix is optimal. For convenience, we still record the state as $\rho$.
\\
(i)\,If $\text{C}^*=\min\{\text{C}_A,\text{C}_B,\text{C}^*\}$, where
$\text{C}^*=2|\rho_{14}+\rho_{23}|$, go to the next step.
\\
(ii)\,If $\text{C}_A=\min\{\text{C}_A,\text{C}_B,\text{C}^*\}$, we set $\Tilde{\rho}=U_{12}\rho\,U_{12}^\dagger$. Calculate the value of $\mathcal{C}(\Tilde{\rho}_A;H_A)+\mathcal{C}(\Tilde{\rho}_B;H_B)$ and record it as $\text{c}_3$. Then go to next step.
\\
(iii)\,If $\text{C}_B=\min\{\text{C}_A,\text{C}_B,\text{C}^*\}$, we set $\Tilde{\rho}=U_{13}\rho\,U_{13}^\dagger$. Calculate the value of $\mathcal{C}(\Tilde{\rho}_A;H_A)+\mathcal{C}(\Tilde{\rho}_B;H_B)$ and record it as $\text{c}_3$. Then go to next step.
\\

{\bf Step 4.} Select the maximum value among $\text{c}_1$, $\text{c}_2$ and $\text{c}_3$. Trace back the optimization path through this value. If $\max\{\text{c}_1,\text{c}_2,\text{c}_3\}>\text{c}_1$, it means that our protocol achieves subsystem capacity gain.

Note that the maximum value is $\text{c}_1$ does not necessarily imply the failure of our protocol. It may be that the residual battery capacity $\triangle(\rho)$ in this state cannot be reduced through such unitary evolution. The specific process of our protocol is shown in Figure 2.
\begin{figure}[htbp]
	\centering
	\includegraphics[width=0.45\textwidth]{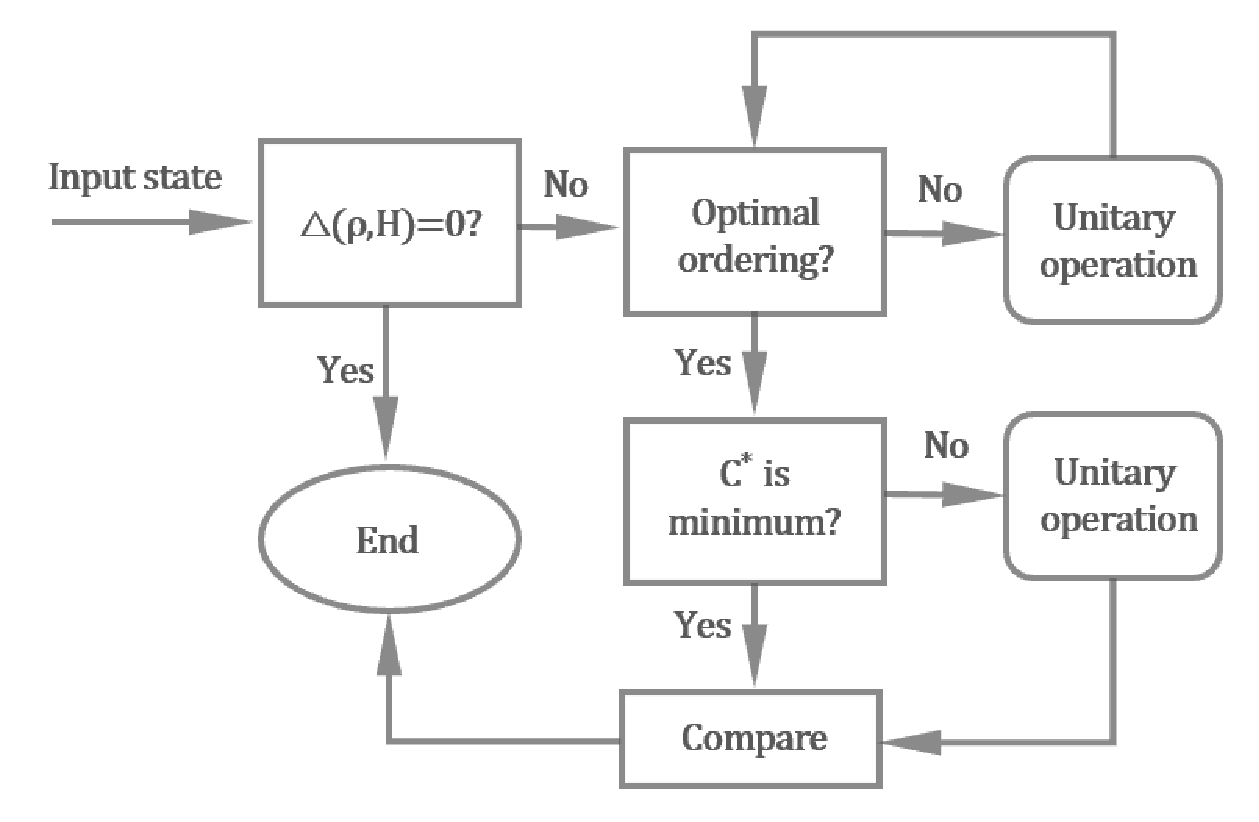}
	\vspace{-1em} \caption{Process of incoherent operation protocol.} \label{Fig.2}
\end{figure}

\mathbi{Example 1.} As an application let us consider the following Bell-like state,
	\begin{equation}\label{exam1}
		\rho_b=\frac{1}{2}\left(
		\begin{array}{cccc}
			1 & 0 & 0 & b\\
			0 & 0 & 0 & 0\\
			0 & 0 & 0 & 0\\
			b & 0 & 0 & 1\\
		\end{array}
		\right ),~~b\in[0,1].
	\end{equation}
	According to the PPT criterion \cite{ap}, this is a family of entangled states whose entanglement increases monotonically with the parameter $b$. We choose the system Hamiltonian $H$ as the longitudinal field XX model with parameters set to $E=\alpha=J=1$. It can be verified that the subsystem battery capacity is zero because its reduced density operators are incoherent states with a uniform population. Hence, the total capacity is entirely residual, i.e. $\mathcal{C}(\rho_b;H)=\triangle(\rho_b,H)=4$. Our incoherent unitary operation protocol shows that we can use $U_{34}$ to achieve subsystem capacity gain. Here, parameter $b$ can be used to quantify the entanglement of the initial state. By calculation, we can obtain that
	\begin{equation*}
		Sub(\Tilde{\rho}_b)=Sub(U_{34}\rho_b U_{34}^\dagger)=2+2b.
	\end{equation*}
	We find that the subsystem battery capacity is proportional to the initial entanglement. At $b=1$, where the initial state $\rho_b$ is maximally entangled, the subsystem capacity gain reaches its maximum. This indicates that incoherent operation can convert all the residual battery capacity into the subsystem battery capacity. We can understand this phenomenon as follows: high entanglement between the two subsystems compresses the subsystem battery capacity because the reduction operation discards a significant amount of quantum information. Our global operation produces a disentangling effect, which drastically reduces this information loss during reduction, thereby enhancing the subsystem capacity.

\mathbi{Example 2.} In this example, we consider a separable, non-X type, single-parameter state,
\begin{equation}\label{exam2}
\rho_a=\frac{1}{6}\left(
\begin{array}{cccc}
	2 & a & 0 & 0\\
	a & 1 & 0 & 0\\
	0 & 0 & 1 & a\\
	0 & 0 & a & 2\\
\end{array}
\right ),
\end{equation}
where $a\in[0,\sqrt{2}]$. We focus on the variations in residual battery capacity $\triangle(\rho_a,H)$ and the subsystem capacity $Sub(\rho_a)$ before and after the incoherent unitary operations for both transverse-field and longitudinal-field Ising models. Our protocol shows that the residual battery capacity $\triangle(\rho_a,H)$ can be suppressed via incoherent operation $U_{34}$, efficiently enhancing the subsystem battery capacity. Figure 3 compares the residual battery capacity $\triangle(\rho_a,H)$ before and after global unitary evolution for interaction strengths of $0.4$, $0.8$ and $1.2$ in both transverse-field and longitudinal-field Ising models, along with the subsystems' capacity changes before and after incoherent operation. We find that in these two models, our incoherent operation compresses the residual battery capacity, thereby enabling the enhancement of the subsystems's capacities, see detailed computations in Appendix D. Additionally, panels (c) and (d) may appear confusing: in the longitudinal-field Ising model, the residual battery capacity remains independent of the spin interaction strength $J$ when $J\leq1$. This behavior stems from the fact that the interaction strength $J$ influences the battery capacity by modifying the eigenvalue ordering of the Hamiltonian. For $J\leq1$, the eigenvalue ordering of the Hamiltonian remains unchanged. However, when $J$ exceeds the critical value of $1$, the eigenvalue ordering abruptly reorganizes, causing the $\triangle(\rho_a,H_I^l)$ to increase with further enhancement of $J$.

Here we are interested in whether the critical value of $J$ depends on specific states or is intrinsically determined by the structural properties of the longitudinal-field Ising Hamiltonian itself. Consider any two-qubit state $\rho$ with eigenvalues ordering $\lambda_3\geq\lambda_2\geq\lambda_1\geq\lambda_0$. The eigenvalues of $H_I^l$ are $\{2E-J, J, J, -2E-J\}$. When $J\leq E$, we have that
\begin{equation*}
	\begin{split}
		\mathcal{C}(\rho;H_I^l)&=(2E-J)(\lambda_3-\lambda_0)+J(\lambda_2-\lambda_1)\\
		&+J(\lambda_1-\lambda_2)+(-2E-J)(\lambda_0-\lambda_3)=4E(\lambda_3-\lambda_0),
	\end{split}
\end{equation*}
which implies that the battery capacity does not depend on the interaction strength $J$ in this scenario. However, when $J>E$, one obtains
\begin{equation*}
	\begin{split}
		\mathcal{C}(\rho;H_I^l)&=J(\lambda_3-\lambda_0)+J(\lambda_2-\lambda_1)\\
		&+(2E-J)(\lambda_1-\lambda_2)+(-2E-J)(\lambda_0-\lambda_3)\\
		&=(2E+2J)(\lambda_3-\lambda_0)+(2J-2E)(\lambda_2-\lambda_1).
	\end{split}
\end{equation*}
Thus in the longitudinal-field Ising model, the critical value of $J$ numerically equals the magnetic field strength $E$, independent of specific quantum states. In this case, $J\approx E$ indicates that the Hamiltonian resides in a strongly competitive regime, transiting from external magnetic field dominance to interaction-driven dynamics.
\begin{figure*}[htbp]
	\centering
	\includegraphics[width=\textwidth]{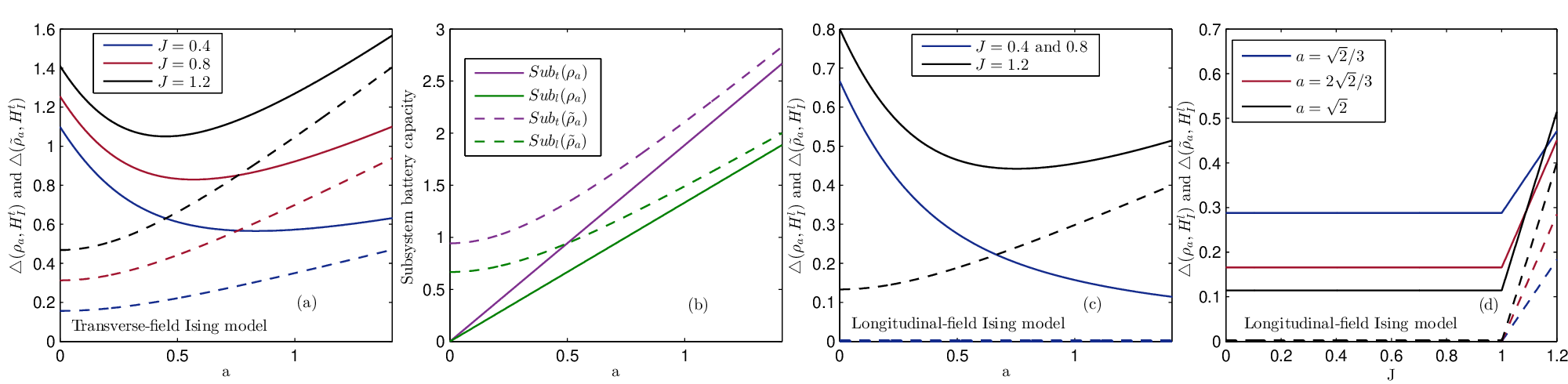}
	\vspace{-1em} \caption{Panel (a) displays the residual battery capacity of the transverse-field Ising model before and after incoherent operation. Panels (c) and (d) shows the corresponding comparison for the longitudinal-field Ising model. The dashed lines represent $\triangle(\tilde{\rho}_a,H)$, while the solid lines represent $\triangle(\rho_a,H)$. Panel (b) presents the changes of the subsystem battery capacity before and after incoherent operations for both transverse-field and longitudinal-field Ising models.} \label{Fig.3}
\end{figure*}
\\

For three-qubit systems, we have the following results parallel to those for two-qubit systems, see Appendix E.

\begin{theorem}
For any three-qubit quantum state $\varrho$, the following trade-off relation holds:
$\mathcal{C}(\varrho_A;H_A)+\mathcal{C}(\varrho_B;H_B)
+\mathcal{C}(\varrho_C;H_C)\leq\mathcal{C}(\varrho;H)$
for any Hamiltonian $H=H_0+H_{int}$ satisfying $\mathcal{C}(\varrho;H)\geq\mathcal{C}(\varrho;H_0)$.
\end{theorem}
In the three-qubit case, the residual battery capacity of a general three-qubit state $\varrho$ is similarly give by
	\begin{equation}\label{e12}
		\triangle(\varrho,H)=\mathcal{C}(\varrho;H)-Sub(\varrho),
	\end{equation}
where $Sub(\varrho)=\mathcal{C}(\varrho_A;H_A)+\mathcal{C}(\varrho_B;H_B)+\mathcal{C}(\varrho_C;H_C)$ is the subsystem capacity. Correspondingly the incoherent and coherent parts of $Sub(\varrho)$ are
	\begin{align}
		Sub_{ic}(\varrho)&=\mathcal{C}(\tau_A;H_A)+\mathcal{C}(\tau_B;H_B)+\mathcal{C}(\tau_C;H_C),\label{e13}\\
		Sub_c(\varrho)&=Sub(\varrho)-Sub_{ic}(\varrho),\label{e14}
	\end{align}
where $\tau_A$, $\tau_B$, and $\tau_C$ are the reduced state of incoherent state $\tau=\text{diag}(\varrho_{11},\varrho_{22},\dots,\varrho_{88})$.

By following an approach analogous to the proof of Theorem 2, we obtain:
\begin{theorem}
	For any three-qubit quantum state $\varrho$, the value of $Sub_{ic}(\varrho)$ can be increased by incoherent unitary operations. In particular, the maximum eigenvalues of $\varrho_A$, $\varrho_B$, $\varrho_C$, $\Tilde{\tau}_A$, $\Tilde{\tau}_B$ and $\tilde{\tau}_C$ are denoted as $\lambda_1^A$, $\lambda_1^B$, $\lambda_1^C$, $\xi_1^A$, $\xi_1^B$, and $\xi_1^C$, respectively. If the incoherent operation $U$ makes 
	\begin{equation}\label{e15}
		\xi_1^A+\xi_1^B+\xi_1^C>\lambda_1^A+\lambda_1^B+\lambda_1^C,
	\end{equation}
	then this process can enhance the subsystem capacity.
\end{theorem}
We still claim that this unitary operation process achieves subsystem capacity gain or subsystem capacity enhancement if 
\begin{equation}\label{e16}
	Sub(\tilde{\varrho})-Sub(\varrho)>0,
\end{equation}
where $\tilde{\varrho}=U\varrho\,U^\dagger$ is the post-evolution state. For three-qubit systems, we present a protocol based on incoherent unitary operation in Appendix F that guides the selection of appropriate operations under certain conditions to achieve subsystem capacity enhancement.

\section{III. The minimal time to achieve subsystem capacity gain} 
We have proposed an incoherent-operation-based protocol to enhance the subsystems' battery capacities, for which the battery states need to undergo at least one quantum gate operation. In this case, the operational evolution time of the quantum battery systems under these gate implementations becomes a practical consideration. We focus on the minimum time required to implement these incoherent operations in the following.

Determining the minimal time required to implement an arbitrary unitary transformation has remained a challenging problem \cite{nkrb,blzh,bzss}. The authors in \cite{nkrb} investigated the problem of steering a system from some initial state to a specified final state by using the Cartan decomposition of unitary operators, with a controllable right-invariant system governed by a Hamiltonian containing a drift term and a local control term. For the two-qubit system, the problem is related to the special unitary group $\mathcal{G}=U(4)$. Since any two-qubit gate can be decomposed as the product of a gate $U\in SU(4)$ and a global phase shift $e^{i\theta}$, the problem reduces to studying the $SU(4)$ group rather than $U(4)$. The Lie algebra of $SU(4)$ has a Cartan decomposition $g=p\oplus l$, where
\begin{equation*}
\begin{split}
l&=\text{span}\frac{i}{2}\{\sigma_x^1,\sigma_y^1,\sigma_z^1,\sigma_x^2,\sigma_y^2,\sigma_z^2\},\\
p&=\text{span}\frac{i}{2}\{\sigma_i^1\sigma_j^2|\,i,j=x,y,z\}.
\end{split}
\end{equation*}
Here $\sigma_x^k$, $\sigma_y^k$, and $\sigma_z^k$ are the standard Pauli matrices acting on the $k$th qubit, $k=1,2$. The Cartan subalgebra is
\begin{equation*}
s=\text{span}\frac{i}{2}\{\sigma_i^1\sigma_i^2|\,i=x,y,z\}.
\end{equation*}
We denote the set of all local gates by $\mathcal{K}$. So $l$ is the Lie subalgebra associated with $\mathcal{K}$. From the Cartan subalgebra $s$, any two-qubit gate $U\in SU(4)$ can be decomposed into a combination of local operations and non-local operations,
\begin{equation}\label{e17}
U=k_1\text{exp}\{\frac{i}{2}(a_1\sigma_x^1\sigma_x^2+a_2\sigma_y^1\sigma_y^2
+a_3\sigma_z^1\sigma_z^2)\}k_2,
\end{equation}
where $a_i\in\mathbb{R}$ $(i=1,2,3)$, $k_1, k_2\in\mathcal{K}$. Khaneja et al. derived the analytical expression for the minimum time required to implement a quantum gate $U$ of the form (\ref{e17}) in heteronuclear spin systems,
\begin{equation*}
t^*=\frac{1}{\pi J}\min\sum_{i=1}^{3}a_i,\,a_i\geq0,
\end{equation*}
where the minimum goes over all possible decompositions in (\ref{e17}).
Since the decomposition given by Eq.\,(\ref{e17}) for a specific $U$ is not unique, determining the minimum time $t^*$ becomes highly challenging.

The authors of Ref.\,\cite{blzh,bzss} further derived an analytical expression for this minimum time by leveraging local invariants. Building on this framework, we investigate the minimum time required to implement a unitary operator within the transverse-field Ising model and XXZ model, as studied in the previous section,
\begin{equation}\label{e18}
\begin{split}
&H_I=J\sigma_z^{\otimes2}+\sum_{i=1}^{4}\nu_iH_i,\\
&H_{xxz}=J[\sigma_z^{\otimes2}+\alpha(\sigma_x^{\otimes2}
+\sigma_y^{\otimes2})]+\sum_{i=1}^{4}\nu_iH_i,
\end{split}
\end{equation}
where $J$ is the coupling strength, $H_1=\sigma_x^1$, $H_2=\sigma_y^1$, $H_3=\sigma_x^2$ and $H_4=\sigma_y^2$, $\alpha$ is the anisotropy parameter satisfying $|\alpha|\leq1$, the last term $\sum_{i=1}^{4}\nu_iH_i$ represents the control Hamiltonian with the coefficients $\nu_i\in\mathbb{R}$ that can be externally manipulated \cite{nkrb}. We have the following results, see the proof in Appendix G.

\begin{theorem}
For the transverse-field Ising model and XXZ model with the system Hamiltonian given by Eq.\,(\ref{e18}), the minimum time required to implement a single incoherent operation under in our protocol is given by
\begin{equation}\label{e19}
\begin{split}
&t^*(U_{14})=t^*(U_{23})=\frac{3\pi}{2J},\\
&t^*(U_{12})=t^*(U_{34})=t^*(U_{13})=t^*(U_{24})=\frac{\pi}{2J},
\end{split}
\end{equation}
where $J$ is the coupling strength given in Eq.\,(\ref{e18}).
\end{theorem}

\section{IV. Discussion and Conclusion}
We have investigated the trade-off relations of battery capacity for general quantum states. For a class of Hamiltonians encompassing Ising, XX, XXZ and XXX models, we have rigorously proved the battery capacity trade-off relations in two-qubit system, demonstrating that the sum of subsystems' battery capacities never exceeds the total system capacity. Furthermore, we have developed an incoherent unitary operation protocol to enhance the subsystem battery capacity, establishing a sufficient condition for achieving the subsystem capacity gain via unitary operations. Our numerical simulations have validated both the rationality of the proposed definitions and the effectiveness of the evolution protocol. Additionally, the results are extended to the case of three-qubit system. Finally, we have determined the minimum time required to enhance the subsystem battery capacity via a single incoherent operation in our protocol. Our results may highlight further studies on battery capacity trade-off relations in higher-dimensional bipartite or multipartite quantum systems under more general evolution schemes that enhance subsystem battery capacities without damaging the total system's battery capacity.

In future work, it would be valuable to investigate the coherent and incoherent battery capacities of different quantum states, as well as a broader range of incoherent operation protocols to enhance subsystem battery capacity. Moreover, like the authors of \cite{mlsl} revealed the impact of entanglement on energy transfer efficiency and stored energy, studying the influence of different quantumness on battery capacity and subsystem capacity gain presents an interesting direction. Finally, applying our theoretical framework and proposed protocols to specific physical systems is the problem of practical relevance.
	
\bigskip
{\bf Acknowledgements:} ~This work is supported by the National Natural Science Foundation of China (NSFC) under Grant Nos.12564048, 12204137 and 12171044; the specific research fund of the Innovation Platform for Academicians of Hainan Province.

\begin{widetext}
\appendix
\section{Appendix A: Proof of Theorem 1}
\setcounter{equation}{0}
\renewcommand{\theequation}{A\arabic{equation}}

We only need to prove that Eq.\,(\ref{e7}) holds for $H_0$ to complete the proof of the Theorem. Eq.\,(\ref{e3}$\sim$\ref{e5}) imply that
\begin{align}
	\lambda_1^A&\leq\lambda_2+\lambda_3,\\
	\lambda_1^B&\leq\lambda_2+\lambda_3,\\
	\lambda_1^A+\lambda_1^B&\leq2\lambda_3+\lambda_1+\lambda_2.
\end{align}
Assuming $\lambda_0^A-\lambda_0^B=c\geq0$ without loss of generality, we have
\begin{equation}
	\begin{split}
		 \lambda_1^A-\lambda_0^A+c&=\lambda_1^A-\lambda_0^B=\lambda_1^A
+\lambda_1^B-1\leq2\lambda_3+\lambda_1+\lambda_2-1=\lambda_3-\lambda_0,
	\end{split}
\end{equation}
where the inequality is derived from Eq.\,(A3). The above inequality implies that
\begin{equation}
	\begin{split}		 \lambda_1^B-\lambda_0^B&=\lambda_1^B-\lambda_0^A+c
=\lambda_1^A+c-\lambda_0^A+c\leq\lambda_3-\lambda_0+c,
	\end{split}
\end{equation}
where the second equality is due to the trace condition of the density operator.
Therefore, we have
\begin{equation*}
	\begin{split}	 \mathcal{C}(\rho_A;H_A)+\mathcal{C}(\rho_B;H_B)&=2\sqrt{2}
E(\lambda_1^A-\lambda_0^A)+2\sqrt{2}E(\lambda_1^B-\lambda_0^B)\leq2\sqrt{2}
E(\lambda_3-\lambda_0-c)+2\sqrt{2}E(\lambda_3-\lambda_0+c)=\mathcal{C}(\rho;H_0),
	\end{split}
\end{equation*}
where the inequality is derived from (A4) and (A5). For the longitudinal field, we get
\begin{equation*}
	\begin{split}
		 \mathcal{C}(\rho_A;H_A)+\mathcal{C}(\rho_B;H_B)&=2E(\lambda_1^A-\lambda_0^A)
+2E(\lambda_1^B-\lambda_0^B)\leq2E(\lambda_3-\lambda_0-c)+2E(\lambda_3-\lambda_0+c)
=\mathcal{C}(\rho;H_0).
	\end{split}
\end{equation*}
Therefore,
\begin{equation}\label{ae}
	\mathcal{C}(\rho_A;H_A)+\mathcal{C}(\rho_B;H_B)\leq\mathcal{C}(\rho;H_0).
\end{equation}
The equality holds in (\ref{ae}) when 
\begin{equation}\label{ae1}
	\begin{split}		 \lambda_3-\lambda_0&=\frac{1}{2}(\lambda_1^A+\lambda_1^B
-\lambda_0^A-\lambda_0^B)=\frac{1}{2}(\lambda_1^A+\lambda_1^B-(1-\lambda_1^A)
-(1-\lambda_1^B))=\lambda_1^A+\lambda_1^B-1.
	\end{split}
\end{equation}

\section{Appendix B: Theorem 1 for the Ising ($J>0$, $\alpha=0$, $\beta=1$), XXZ ($J>0$, $\alpha\neq0$, $\beta=1$), XX ($J>0$, $\alpha\neq0$, $\beta=0$) and XXX models ($J>0$, $\alpha=\beta=1$)}
\setcounter{equation}{0}
\renewcommand{\theequation}{B\arabic{equation}}
According to Theorem 1, we only demonstrate that for a two-qubit quantum state, the battery capacity corresponding to the interaction-free Hamiltonian does not exceed that of the transverse-field Ising model, transverse-field XXZ model and transverse-field XXX model.

Given a two-qubit state $\rho$ with eigenvalues $\lambda_0\leq\lambda_1\leq\lambda_2\leq\lambda_3$. For the non-interaction Hamiltonian $H_0=E(\sigma_x\otimes I_2+\sigma_y\otimes I_2+I_2\otimes\sigma_x+I_2\otimes\sigma_y)$ and the Ising model Hamiltonian $H_I=H_0+J\sigma_z\otimes\sigma_z$, one has
\begin{equation*}
	\begin{split} \mathcal{C}(\rho;H_0)&=4\sqrt{2}E(\lambda_3-\lambda_0)
\leq[\sqrt{8E^2+J^2}-(-\sqrt{8E^2+J^2})](\lambda_3-\lambda_0)\leq[\sqrt{8E^2+J^2}
-(-\sqrt{8E^2+J^2})](\lambda_3-\lambda_0)\\		 &+(J-(-J))(\lambda_2-\lambda_1)\leq(\eta_3-\eta_0)(\lambda_3-\lambda_0)
+(\eta_2-\eta_1)(\lambda_2-\lambda_1)=\sum_{i=0}^{3}\lambda_i(\eta_i-\eta_{3-i})
=\mathcal{C}(\rho;H_I),
	\end{split}
\end{equation*}
where $\{\eta_i\}$ is the eigenvalues $\{\pm J, \pm\sqrt{8E^2+J^2}\}$ of $H_I$ in ascending order. 

For the XXZ model $H_{xxz}=H_0+J(\sigma_z^{\otimes2}+\alpha(\sigma_x^{\otimes2}+\sigma_y^{\otimes2}))$ with eigenvalues $\{J, -J-2\alpha J, \alpha J\pm\sqrt{8E^2+J^2(\alpha-1)^2}\}\equiv\{\epsilon_0,\epsilon_1,\epsilon_2,\epsilon_3\}$ in ascending order, we have
\begin{equation*}
	\begin{split}
		\mathcal{C}(\rho;H_0)&=4\sqrt{2}E(\lambda_3-\lambda_0)\leq[(\alpha J+\sqrt{8E^2+J^2(\alpha-1)^2})-(\alpha J-\sqrt{8E^2+J^2(\alpha-1)^2})](\lambda_3-\lambda_0)\\
		 &\leq(\epsilon_3-\epsilon_0)(\lambda_3-\lambda_0)+(\epsilon_2-\epsilon_1)
(\lambda_2-\lambda_1)=\sum_{i=0}^{3}\lambda_i(\epsilon_i-\epsilon_{3-i})
=\mathcal{C}(\rho;H_{xxz}).
	\end{split}
\end{equation*}

For the XX model $H_{xx}=H_0+\alpha J(\sigma_x^{\otimes2}+\sigma_y^{\otimes2})$ with eigenvalues $\{0, -2\alpha J, \alpha J\pm\sqrt{8E^2+\alpha^2J^2}\}\equiv\{\iota_0,\iota_1,\iota_2,\iota_3\}$ in ascending order, we have
\begin{equation*}
	\begin{split}
		\mathcal{C}(\rho;H_0)&=4\sqrt{2}E(\lambda_3-\lambda_0)\leq[(\alpha J+\sqrt{8E^2+\alpha^2J^2})-(\alpha J-\sqrt{8E^2+\alpha^2J^2})](\lambda_3-\lambda_0)\\
		&\leq(\iota_3-\iota_0)(\lambda_3-\lambda_0)+(\iota_2-\iota_1)
		(\lambda_2-\lambda_1)=\sum_{i=0}^{3}\lambda_i(\iota_i-\iota_{3-i})
		=\mathcal{C}(\rho;H_{xx}).
	\end{split}
\end{equation*}

For the case of longitudinal external magnetic, we have
\begin{equation*}
	\begin{split}		 \mathcal{C}(\rho;H_0)&=4E(\lambda_3-\lambda_0)
=[(2E-J)-(-2E-J)](\lambda_3-\lambda_0)=[(2E-J)-(-2E-J)]
(\lambda_3-\lambda_0)+(J-J)(\lambda_2-\lambda_1)\\		 &\leq(\eta_3-\eta_0)(\lambda_3-\lambda_0)+(\eta_2-\eta_1)
(\lambda_2-\lambda_1)=\sum_{i=0}^{3}\lambda_i(\eta_i-\eta_{3-i})=\mathcal{C}(\rho;H_I),
	\end{split}
\end{equation*}
where $\{\eta_i\}$ denotes the eigenvalues $\{\pm 2E-J, J,J\}$ of $H_I$ in ascending order. 

For the XXZ model $H_{xxz}$ with eigenvalues $\{\pm 2E-J, \pm 2\alpha J+J\}\equiv\{\epsilon_0,\epsilon_1,\epsilon_2,\epsilon_3\}$ in ascending order, we have
\begin{equation*}
	\begin{split}		 \mathcal{C}(\rho;H_0)&=4E(\lambda_3-\lambda_0)=[(2E-J)-(-2E-J)]
(\lambda_3-\lambda_0)<(\epsilon_3-\epsilon_0)(\lambda_3-\lambda_0)
+(\epsilon_2-\epsilon_1)(\lambda_2-\lambda_1)\\
		&=\sum_{i=0}^{3}\lambda_i(\epsilon_i-\epsilon_{3-i})=\mathcal{C}(\rho;H_{xxz}).
	\end{split}
\end{equation*}

For the XX model $H_{xx}$ with eigenvalues $\{\pm 2E, \pm 2\alpha J\}\equiv\{\iota_0,\iota_1,\iota_2,\iota_3\}$ in ascending order, we have
\begin{equation*}
	\begin{split}		 \mathcal{C}(\rho;H_0)&=4E(\lambda_3-\lambda_0)=[2E-(-2E)]
		(\lambda_3-\lambda_0)<(\iota_3-\iota_0)(\lambda_3-\lambda_0)
		+(\iota_2-\iota_1)(\lambda_2-\lambda_1)\\
		&=\sum_{i=0}^{3}\lambda_i(\iota_i-\iota_{3-i})=\mathcal{C}(\rho;H_{xx}).
	\end{split}
\end{equation*}
In particular, taking $\alpha=1$ in the aforementioned XXZ case yields $\mathcal{C}(\rho;H_0)\leq\mathcal{C}(\rho;H_{xxx})$.

\section{Appendix C: Proof of Theorem 2}
\setcounter{equation}{0}
\renewcommand{\theequation}{C\arabic{equation}}
Theorem 2 is in fact independent of the choice of Hamiltonian. To see this point, we first investigate the relationship between the spectrum of the non-interaction Hamiltonian $H_0$ in general case and those of its corresponding subsystems $H_A$ and $H_B$. The general form of $H_0$ can be written as,
\begin{equation*}
	H_0=E_1(\sigma_x\otimes I_2+I_2\otimes\sigma_x)+E_2(\sigma_y\otimes I_2+I_2\otimes\sigma_y)+E_3(\sigma_z\otimes I_2+I_2\otimes\sigma_z),
\end{equation*}
where $E_1$, $E_2$ and $E_3$ correspond to the magnetic field strengths along the $x$, $y$ and $z$ directions, respectively. Then the corresponding subsystems' Hamiltonians are
$H_A=H_B=E_1\sigma_x+E_2\sigma_y+E_3\sigma_z$. Hence, the spectra of the total system Hamiltonian and the subsystems' Hamiltonian are $\text{Spec}(H_0)=\{\pm2\sqrt{E_1^2+E_2^2+E_3^2}, 0, 0\}$ and $\text{Spec}(H_A)=\text{Spec}(H_B)=\{\pm\sqrt{E_1^2+E_2^2+E_3^2}\}$.
Let
$$
H=\sum_{i=1}^{4}\varepsilon_i|\varepsilon_i\rangle\langle\varepsilon_i|,\,\,\,\,
H_A=\sum_{i=1}^{2}\varepsilon_i^A|\varepsilon_i^A\rangle\langle\varepsilon_i^A|,,\,\,\,\,
H_B=\sum_{i=1}^{2}\varepsilon_i^B|\varepsilon_i^B\rangle\langle\varepsilon_i^B|
$$
be the spectral decompositions of the total system Hamiltonian and the subsystems' Hamiltonians, with their eigenvalues ordered in ascending sequence. It follows that $\varepsilon_4=2\varepsilon_2^A=2\varepsilon_2^B$ and $\varepsilon_1=2\varepsilon_1^A=2\varepsilon_1^B$.

Given a two-qubit state $\rho=(\rho_{ij})$, its incoherent state is given by $\tau=\text{diag}(\rho_{11},\rho_{22},\rho_{33},\rho_{44})$. We arrange these diagonal elements in ascending order and denote them as $\{\mu_i\}_{i=1}^4$. Mathematically, it is easy to verify that there is a unitary matrix $U$ (possibly the product of a series of unitary matrices) such that $\Tilde{\tau}=U\tau\,U^\dagger=\text{diag}(\mu_4,\mu_3,\mu_2,\mu_1)$. Hence, we have
\begin{equation*}
\begin{split}
Sub_{ic}(\tilde{\rho})-Sub_{ic}(\rho)&=[\mathcal{C}(\Tilde{\tau}_A;H_A)
+\mathcal{C}(\Tilde{\tau}_B;H_B)]-[\mathcal{C}(\tau_A;H_A)+\mathcal{C}(\tau_B;H_B)]\\
&=(\varepsilon_2^A-\varepsilon_1^A)(\mu_{44}+\mu_{33}-\mu_{22}-\mu_{11})
+(\varepsilon_2^B-\varepsilon_1^B)(\mu_{44}+\mu_{22}-\mu_{33}-\mu_{11})\\
&-(\varepsilon_2^A-\varepsilon_1^A)|\rho_{11}+\rho_{22}-\rho_{33}-\rho_{44}|
-(\varepsilon_2^B-\varepsilon_1^B)|\rho_{11}+\rho_{33}-\rho_{22}-\rho_{44}|\\
&\geq0.
\end{split}
\end{equation*}
This means that $Sub_{ic}(\rho)$ is enhanced by our scheme.

Let us now consider the conditions under which this process guarantees the subsystem capacity gain. The diagonal elements of every positive semi-definite matrix are majorized by eigenvalues \cite{wyd}, which implies that
\begin{equation}\label{b}
	\mathcal{C}(\Tilde{\rho}_A;H_A)\geq\mathcal{C}(\Tilde{\tau}_A;H_A),\, \mathcal{C}(\Tilde{\rho}_B;H_B)\geq\mathcal{C}(\Tilde{\tau}_B;H_B),
\end{equation}
where $\Tilde{\tau}_A$ and $\Tilde{\tau}_B$ are the incoherent state of $\Tilde{\rho}_A$ and $\Tilde{\rho}_B$, respectively, and we have used the fact that the capacity $\mathcal{C}$ is a Schur-convex functional. (\ref{b}) actually provides us with a sufficient condition for achieving subsystem capacity gain,
\begin{equation*}
	\begin{split} &\mathcal{C}(\rho_A;H_A)+\mathcal{C}(\rho_B;H_B)
=(\varepsilon_2^A-\varepsilon_1^A)(\lambda_1^A-\lambda_0^A)
+(\varepsilon_2^B-\varepsilon_1^B)(\lambda_1^B-\lambda_0^B)
=2(\varepsilon_2^A-\varepsilon_1^A)(\lambda_1^A+\lambda_1^B-\lambda_0^A-\lambda_0^B)\\ &<2(\varepsilon_2^A-\varepsilon_1^A)(\xi_1^A+\xi_1^B-\xi_0^A-\xi_0^B)
=(\varepsilon_2^A-\varepsilon_1^A)(\xi_1^A-\xi_0^A)
+(\varepsilon_2^B-\varepsilon_1^B)(\xi_1^B-\xi_0^B)
=\mathcal{C}(\Tilde{\tau}_A;H_A)+\mathcal{C}(\Tilde{\tau}_B;H_B)\\
		&\leq\mathcal{C}(\Tilde{\rho}_A;H_A)+\mathcal{C}(\Tilde{\rho}_B;H_B),
	\end{split}
\end{equation*}
which completes the proof.

\section{Appendix D: State (\ref{exam}) for the transverse and longitudinal-field Ising models}
\setcounter{equation}{0}
\renewcommand{\theequation}{D\arabic{equation}}
The eigenvalues of $\rho_a$ are
\begin{equation}
	\begin{split}
		\lambda_0=\lambda_1&=\frac{1}{4}-\frac{1}{12}\sqrt{4a^2+1},\,\,\,\,
		\lambda_2=\lambda_3=\frac{1}{4}+\frac{1}{12}\sqrt{4a^2+1}.
	\end{split}
\end{equation}
Corresponding to transverse-field Ising model and longitudinal-field Ising model, the whole system Hamiltonians are
\begin{align}
	&H_I^t=E(\sigma_x\otimes I_2+\sigma_y\otimes I_2+I_2\otimes\sigma_x+I_2\otimes\sigma_y)-J\sigma_z\otimes\sigma_z\otimes\sigma_z,\nonumber\\
	&H_I^l=E(\sigma_z\otimes I_2+I_2\otimes\sigma_z)-\sigma_z\otimes\sigma_z\otimes\sigma_z,\nonumber
\end{align}
respectively. According to (\ref{e1}) and (E1), the quantum battery capacity under these two models are $\mathcal{C}(\rho_a;H_I^t)=\frac{1}{3}\sqrt{4a^2+1}(\sqrt{8+J}+J)$, and
$$
\mathcal{C}(\rho_a;H_I^l)=\left\{\begin{array}{rcl}
	\frac{2}{3}\sqrt{4a^2+1},& J\in[0,1]\\
	\\
	\frac{2J}{3}\sqrt{4a^2+1}.& J\in(1,1.2]
\end{array}\right.
$$
As the reduced states of $\rho_a$ are
$$\rho_a^A=\frac{1}{2}\left(
\begin{array}{cc}
	1 & 0 \\
	0 & 1 \\
\end{array}	
\right),~~\rho_a^B=\frac{1}{2}\left(
\begin{array}{cc}
	1 & \frac{2a}{3} \\
	\frac{2a}{3} & 1 \\
\end{array}	
\right),$$
one has $\mathcal{C}(\rho_a^A;H_A)=0$ and $\mathcal{C}(\rho_a^B;H_B)=4\sqrt{2}a/3$.
From (\ref{e8}), the residual battery capacity $\triangle(\rho,H)$ with respect to these two models are $\triangle(\rho_a,H_I^t)=\frac{1}{3}\sqrt{4a^2+1}(\sqrt{8+J}+J)-4\sqrt{2}a/3$, and
$$
\triangle(\rho_a,H_I^l)=\left\{\begin{array}{rcl}
	&\frac{2}{3}\sqrt{4a^2+1}-\frac{4a}{3},~\,J\in[0,1]~~~\\
	\\
	&\frac{2J}{3}\sqrt{4a^2+1}-\frac{4a}{3}.~\,J\in(1,1.2]
\end{array}\right.
$$
In the above calculations, we have employed the subsystem battery capacity $Sub_l(\rho_a)=4a/3$ within the longitudinal-field Ising model.

The incoherent operation protocol shows that we can select the unitary operation $U_{34}$, i.e.,
$$\Tilde{\rho}_a=U_{34}\rho_a\,U_{34}^\dagger=\frac{1}{6}\left(
\begin{array}{cccc}
	2 & a & 0 & 0\\
	a & 1 & 0 & 0\\
	0 & 0 & 2 & a\\
	0 & 0 & a & 1\\
\end{array}
\right ).$$
Then the reduced states of $\Tilde{\rho}_a$ are
$$\Tilde{\rho}_a^A=\frac{1}{2}\left(
\begin{array}{cc}
	1 & 0 \\
	0 & 1 \\
\end{array}	
\right),~~\Tilde{\rho}_a^B=\frac{1}{3}\left(
\begin{array}{cc}
	2 & a \\
	a & 1 \\
\end{array}	
\right).
$$
After unitary operation, the subsystem battery capacity can be written as
\begin{equation}
\begin{split}
Sub_t(\tilde{\rho_a})&=\frac{2\sqrt{2}}{3}\sqrt{4a^2+1},\,\,\,\,Sub_l(\tilde{\rho_a})=\frac{2}{3}\sqrt{4a^2+1}.
\end{split}
\end{equation}
The residual battery capacity for these two models are $\triangle(\Tilde{\rho}_a,H_I^t)=\frac{1}{3}\sqrt{4a^2+1}(\sqrt{8+J}+J)
-\frac{2\sqrt{2}}{3}\sqrt{4a^2+1}$,
and
$$
\triangle(\Tilde{\rho}_a,H_I^l)=\left\{\begin{array}{rcl}
	&0,~\,J\in[0,1]~~~~~~~~~~~~~~~~~~~~~~~~~~~~~~~~\\
	\\
	&\frac{2J}{3}\sqrt{4a^2+1}-\frac{2}{3}\sqrt{4a^2+1}.~\,J\in(1,1.2]
\end{array}\right.
$$

\section{Appendix E: Proof of Theorem 3}
\setcounter{equation}{0}
\renewcommand{\theequation}{E\arabic{equation}}
Similar to the two-qubit case, we also consider two types of external magnetic field terms:
\begin{equation*}
	\begin{split}
		H_0&=E(\sigma_z\otimes I_2\otimes I_2+I_2\otimes\sigma_z\otimes I_2+I_2\otimes I_2\otimes\sigma_z),\\
		H_0&=E(\sigma_x\otimes I_2\otimes I_2+I_2\otimes\sigma_x\otimes I_2+I_2\otimes I_2\otimes\sigma_x+\sigma_y\otimes I_2\otimes I_2+I_2\otimes\sigma_y\otimes I_2+I_2\otimes I_2\otimes\sigma_y).
	\end{split}
\end{equation*}
We only need to prove
\begin{equation} \mathcal{C}(\varrho_A;H_A)+\mathcal{C}(\varrho_B;H_B)+\mathcal{C}(\varrho_C;H_C)\leq\mathcal{C}(\varrho;H_0).
\end{equation}
The solution of three-qubit QMP demonstrates that the reduced states $\varrho_A, \varrho_B, \varrho_C$, with spectra $\{\lambda_1^A\leq\lambda_2^A\}, \{\lambda_1^B\leq\lambda_2^B\}, \{\lambda_1^C\leq\lambda_2^C\}$, respectively, serve as marginals of a global state $\varrho$ with spectrum $\{\lambda_1\leq\dots\leq\lambda_8\}$ only if the following inequalities hold \cite{aak}:
\begin{align}
&\triangle_3\leq\lambda_8+\lambda_7+\lambda_6+\lambda_5-\lambda_4-\lambda_3-\lambda_2-\lambda_1,\nonumber\\
	&\triangle_2+\triangle_3\leq2\lambda_8+2\lambda_7-2\lambda_2-2\lambda_1,\nonumber\\ &\triangle_1+\triangle_2+\triangle_3\leq3\lambda_8+\lambda_7+\lambda_6+\lambda_5-\lambda_4-\lambda_3-\lambda_2-3\lambda_1,\nonumber\\
	 -&\triangle_1+\triangle_2+\triangle_3\leq\lambda_8+3\lambda_7+\lambda_6+\lambda_5-\lambda_4-\lambda_3-\lambda_2-3\lambda_1,\nonumber\\
	 -&\triangle_1+\triangle_2+\triangle_3\leq3\lambda_8+\lambda_7+\lambda_6+\lambda_5-\lambda_4-\lambda_3-3\lambda_2-\lambda_1,\nonumber\\
	 &\triangle_1+\triangle_2+2\triangle_3\leq4\lambda_8+2\lambda_7+2\lambda_6-2\lambda_3-2\lambda_2-4\lambda_1,\nonumber\\
	 -&\triangle_1+\triangle_2+2\triangle_3\leq2\lambda_8+4\lambda_7+2\lambda_6-2\lambda_3-2\lambda_2-4\lambda_1,\nonumber\\
	 -&\triangle_1+\triangle_2+2\triangle_3\leq4\lambda_8+2\lambda_7+2\lambda_5-2\lambda_3-2\lambda_2-4\lambda_1,\nonumber\\
	 -&\triangle_1+\triangle_2+2\triangle_3\leq4\lambda_8+2\lambda_7+2\lambda_6-2\lambda_4-2\lambda_2-4\lambda_1,\nonumber\\
	 -&\triangle_1+\triangle_2+2\triangle_3\leq4\lambda_8+2\lambda_7+2\lambda_6-2\lambda_3-4\lambda_2-2\lambda_1,\nonumber
\end{align}
where $\triangle_1\leq\triangle_2\leq\triangle_3$ denotes $\lambda_2^X-\lambda_1^X$ $(X=A,B,C)$ in ascending order. Therefore, for the case of longitudinal-field we have
\begin{equation*}
	\begin{split}
		 \mathcal{C}(\varrho;H_0)&=6E(\lambda_8-\lambda_1)+2E(\lambda_7-\lambda_2)+2E(\lambda_6-\lambda_3)+2E(\lambda_5-\lambda_4)\\
		 &=2E(3\lambda_8+\lambda_7+\lambda_6+\lambda_5-\lambda_4-\lambda_3-\lambda_2-3\lambda_1)\\
		&\geq2E(\triangle_1+\triangle_2+\triangle_3)\\
		 &=2E(\lambda_2^A-\lambda_1^A)+2E(\lambda_2^B-\lambda_1^B)+2E(\lambda_2^C-\lambda_1^C)\\
		&=\mathcal{C}(\varrho_A;H_A)+\mathcal{C}(\varrho_B;H_B)+\mathcal{C}(\varrho_C;H_C).
	\end{split}
\end{equation*}
Herein, the inequality follows from the third inequality in the solution of the three-qubit QMP. Similarly, when $H_0$ is the transverse-field, one has
\begin{equation*}
	\begin{split}
		 \mathcal{C}(\varrho;H_0)&=6\sqrt{2}E(\lambda_8-\lambda_1)+2\sqrt{2}E(\lambda_7-\lambda_2)+2\sqrt{2}E(\lambda_6-\lambda_3)+2\sqrt{2}E(\lambda_5-\lambda_4)\\
		 &=2\sqrt{2}E(3\lambda_8+\lambda_7+\lambda_6+\lambda_5-\lambda_4-\lambda_3-\lambda_2-3\lambda_1)\\
		&\geq2\sqrt{2}E(\triangle_1+\triangle_2+\triangle_3)\\
		 &=2\sqrt{2}E(\lambda_2^A-\lambda_1^A)+2\sqrt{2}E(\lambda_2^B-\lambda_1^B)+2\sqrt{2}E(\lambda_2^C-\lambda_1^C)\\
		&=\mathcal{C}(\varrho_A;H_A)+\mathcal{C}(\varrho_B;H_B)+\mathcal{C}(\varrho_C;H_C).
	\end{split}
\end{equation*}

\section{Appendix F: Incoherent operation protocol in three-qubit systems}
\setcounter{equation}{0}
\renewcommand{\theequation}{F\arabic{equation}}
For a three-qubit state $\varrho=(\varrho_{ij})_{8\times8}$, we set the energy levels of $H_A$, $H_B$ and $H_C$ as $\varepsilon_0^X\leq\varepsilon_1^X$ $(X=A,B,C)$. Then the battery capacity corresponding to subsystems $A$, $B$ and $C$ are, respectively,
\begin{equation*}
	\begin{split}
		 \mathcal{C}(\varrho_A;H_A)&=(\varepsilon_1^A-\varepsilon_0^A)\sqrt{4|\varrho_{15}+\varrho_{26}+\varrho_{37}+\varrho_{48}|^2+(\varrho_{11}+\varrho_{22}+\varrho_{33}+\varrho_{44}-\varrho_{55}-\varrho_{66}-\varrho_{77}-\varrho_{88})^2},\\
		 \mathcal{C}(\varrho_B;H_B)&=(\varepsilon_1^B-\varepsilon_0^B)\sqrt{4|\varrho_{13}+\varrho_{24}+\varrho_{57}+\varrho_{68}|^2+(\varrho_{11}+\varrho_{22}+\varrho_{55}+\varrho_{66}-\varrho_{33}-\varrho_{44}-\varrho_{77}-\varrho_{88})^2},\\
		 \mathcal{C}(\varrho_C;H_C)&=(\varepsilon_1^C-\varepsilon_0^C)\sqrt{4|\varrho_{12}+\varrho_{34}+\varrho_{56}+\varrho_{78}|^2+(\varrho_{11}+\varrho_{33}+\varrho_{55}+\varrho_{77}-\varrho_{22}-\varrho_{44}-\varrho_{66}-\varrho_{88})^2}.
	\end{split}
\end{equation*}
Analogous to the two-qubit protocol, the subsystem battery capacity can be reformulated in terms of the symbols \text{C} and \text{IC} as
\begin{equation*}
\begin{split}
\mathcal{C}(\varrho_A;H_A)&=(\varepsilon_1^A-\varepsilon_0^A)\sqrt{\text{C}_A^2+\text{IC}_A^2},\,\,
\mathcal{C}(\varrho_B;H_B)=(\varepsilon_1^B-\varepsilon_0^B)\sqrt{\text{C}_B^2+\text{IC}_B^2},\,\,
\mathcal{C}(\varrho_C;H_C)=(\varepsilon_1^C-\varepsilon_0^C)\sqrt{\text{C}_C^2+\text{IC}_C^2}.
\end{split}
\end{equation*}

We introduce several key incoherent unitary operations employed in the protocol below.

(1)\,$U_{12},\,U_{34},\,U_{56},\,U_{78}$. Keep $\text{IC}_A$ and $\text{IC}_B$ unchanged and change the values of $\text{C}_A$, $\text{C}_B$ and $\text{C}_C$.

(2)\,$U_{13},\,U_{24},\,U_{57},\,U_{68}$. Keep $\text{IC}_A$ and $\text{IC}_C$ unchanged and change the values of $\text{C}_A$, $\text{C}_B$ and $\text{C}_C$.

(3)\,$U_{15},\,U_{26},\,U_{37},\,U_{48}$. Keep $\text{IC}_B$ and $\text{IC}_C$ unchanged and change the values of $\text{C}_A$, $\text{C}_B$ and $\text{C}_C$.

We tabulate the variations of $\text{C}_A$, $\text{C}_B$ and $\text{C}_C$ under these twelve unitary evolutions, see Table I.
\begin{table*}[htbp]
	\begin{tabular*}{\textwidth}{@{}@{\extracolsep{\fill}}lllllllllllll@{}}
		\hline
		\hline
		~~   &~~~~~~~~~~~~$\text{C}_A$   &~~~~~~~~~~~~$\text{C}_B$   &~~~~~~~~~~~~$\text{C}_C$ \\
		\hline
		$U_{12}$   &$2|\varrho_{25}+\varrho_{16}+\varrho_{37}+\varrho_{48}|$   &$2|\varrho_{23}+\varrho_{14}+\varrho_{57}+\varrho_{68}|$   &$2|\varrho_{12}^*+\varrho_{34}+\varrho_{56}+\varrho_{78}|$  \\
		\hline
		$U_{34}$   &$2|\varrho_{15}+\varrho_{26}+\varrho_{47}+\varrho_{38}|$   &$2|\varrho_{14}+\varrho_{23}+\varrho_{57}+\varrho_{68}|$   &$2|\varrho_{12}+\varrho_{34}^*+\varrho_{56}+\varrho_{78}|$  \\
		\hline
		$U_{56}$   &$2|\varrho_{16}+\varrho_{25}+\varrho_{37}+\varrho_{48}|$   &$2|\varrho_{13}+\varrho_{24}+\varrho_{67}+\varrho_{58}|$   &$2|\varrho_{12}+\varrho_{34}+\varrho_{56}^*+\varrho_{78}|$  \\
		\hline
		$U_{78}$   &$2|\varrho_{15}+\varrho_{26}+\varrho_{38}+\varrho_{47}|$   &$2|\varrho_{13}+\varrho_{24}+\varrho_{58}+\varrho_{67}|$   &$2|\varrho_{12}+\varrho_{34}+\varrho_{56}+\varrho_{78}^*|$  \\
		\hline
		$U_{13}$   &$2|\varrho_{35}+\varrho_{26}+\varrho_{17}+\varrho_{48}|$   &$2|\varrho_{13}^*+\varrho_{24}+\varrho_{57}+\varrho_{68}|$   &$2|\varrho_{23}^*+\varrho_{14}+\varrho_{56}+\varrho_{78}|$  \\
		\hline
		$U_{24}$   &$2|\varrho_{15}+\varrho_{46}+\varrho_{37}+\varrho_{28}|$   &$2|\varrho_{13}+\varrho_{24}^*+\varrho_{57}+\varrho_{68}|$   &$2|\varrho_{14}+\varrho_{23}^*+\varrho_{56}+\varrho_{78}|$  \\
		\hline
		$U_{57}$   &$2|\varrho_{17}+\varrho_{26}+\varrho_{35}+\varrho_{48}|$   &$2|\varrho_{13}+\varrho_{24}+\varrho_{57}^*+\varrho_{68}|$   &$2|\varrho_{12}+\varrho_{34}+\varrho_{67}^*+\varrho_{58}|$  \\
		\hline
		$U_{68}$   &$2|\varrho_{15}+\varrho_{28}+\varrho_{37}+\varrho_{46}|$   &$2|\varrho_{13}+\varrho_{24}+\varrho_{57}+\varrho_{68}^*|$   &$2|\varrho_{12}+\varrho_{34}+\varrho_{58}+\varrho_{67}^*|$  \\
		\hline
		$U_{15}$   &$2|\varrho_{15}^*+\varrho_{26}+\varrho_{37}+\varrho_{48}|$   &$2|\varrho_{35}^*+\varrho_{24}+\varrho_{17}+\varrho_{68}|$   &$2|\varrho_{25}^*+\varrho_{34}+\varrho_{16}+\varrho_{78}|$  \\
		\hline
		$U_{26}$   &$2|\varrho_{15}+\varrho_{26}^*+\varrho_{37}+\varrho_{48}|$   &$2|\varrho_{13}+\varrho_{46}^*+\varrho_{57}+\varrho_{28}|$   &$2|\varrho_{16}+\varrho_{34}+\varrho_{25}^*+\varrho_{78}|$  \\
		\hline
		$U_{37}$   &$2|\varrho_{15}+\varrho_{26}+\varrho_{37}^*+\varrho_{48}|$   &$2|\varrho_{17}+\varrho_{24}+\varrho_{35}^*+\varrho_{68}|$   &$2|\varrho_{12}+\varrho_{47}^*+\varrho_{56}+\varrho_{38}|$  \\
		\hline
		$U_{48}$   &$2|\varrho_{15}+\varrho_{26}+\varrho_{37}+\varrho_{48}^*|$   &$2|\varrho_{13}+\varrho_{28}+\varrho_{57}+\varrho_{46}^*|$   &$2|\varrho_{12}+\varrho_{38}+\varrho_{56}+\varrho_{47}^*|$  \\
		\hline
		\hline
	\end{tabular*}
		\caption{After applying these twelve unitary evolutions to the initial state, the values of $\text{C}_A$, $\text{C}_B$, and $\text{C}_C$ are as shown in the table.}
\end{table*}

The subsystem battery capacity is influenced by the incoherent component (\text{IC}). The optimal ordering of the two-qubit system with respect to the \text{IC} is given by (D4) and (D5). However, the optimal ordering for the three-qubit case is significantly more complex. Assume the diagonal elements are arranged in ascending order as $\mu_1\leq\dots\leq\mu_8$. The optimal ordering refers to the ordering that satisfies $Sub_{ic}(\varrho)=\mathcal{C}(\tau;H_0)$. According to (\ref{e13}), this requires $\mu_8+\mu_7+\mu_6+\mu_5-\mu_4-\mu_3-\mu_2-\mu_1$, $\mu_8+\mu_7+\mu_3+\mu_4-\mu_5-\mu_6-\mu_2-\mu_1$ and $\mu_8+\mu_6+\mu_4+\mu_2-\mu_1-\mu_3-\mu_5-\mu_7$ corresponding to $\text{IC}_A$, $\text{IC}_B$ and $\text{IC}_C$, respectively. The optimal ordering satisfying $\text{IC}_A\geq\text{IC}_B\geq\text{IC}_C$ is labeled \textbf{optimal ordering A-B-C} in the following, with analogous notations for the other five possible optimal orderings.

{\bf Optimal ordering A-B-C:}
\begin{equation}
\begin{split}
&\varrho_{11}\geq\varrho_{22}\geq\varrho_{33}\geq\varrho_{44}\geq\varrho_{55}\geq\varrho_{66}\geq\varrho_{77}\geq\varrho_{88},\,\,\,\,
\varrho_{22}\geq\varrho_{11}\geq\varrho_{44}\geq\varrho_{33}\geq\varrho_{66}\geq\varrho_{55}\geq\varrho_{88}\geq\varrho_{77},\\
&\varrho_{33}\geq\varrho_{44}\geq\varrho_{11}\geq\varrho_{22}\geq\varrho_{77}\geq\varrho_{88}\geq\varrho_{55}\geq\varrho_{66},\,\,\,\,
\varrho_{44}\geq\varrho_{33}\geq\varrho_{22}\geq\varrho_{11}\geq\varrho_{88}\geq\varrho_{77}\geq\varrho_{66}\geq\varrho_{55},\\
&\varrho_{55}\geq\varrho_{66}\geq\varrho_{77}\geq\varrho_{88}\geq\varrho_{11}\geq\varrho_{22}\geq\varrho_{33}\geq\varrho_{44},\,\,\,\,
\varrho_{66}\geq\varrho_{55}\geq\varrho_{88}\geq\varrho_{77}\geq\varrho_{22}\geq\varrho_{11}\geq\varrho_{44}\geq\varrho_{33},\\
&\varrho_{77}\geq\varrho_{88}\geq\varrho_{55}\geq\varrho_{66}\geq\varrho_{33}\geq\varrho_{44}\geq\varrho_{11}\geq\varrho_{22},\,\,\,\,
\varrho_{88}\geq\varrho_{77}\geq\varrho_{66}\geq\varrho_{55}\geq\varrho_{44}\geq\varrho_{33}\geq\varrho_{22}\geq\varrho_{11}.
\end{split}
\end{equation}

{\bf Optimal ordering A-C-B:}
\begin{equation}
	\begin{split}
		 &\varrho_{11}\geq\varrho_{33}\geq\varrho_{22}\geq\varrho_{44}\geq\varrho_{55}\geq\varrho_{77}\geq\varrho_{66}\geq\varrho_{88},\,\,\,\,
		 \varrho_{22}\geq\varrho_{44}\geq\varrho_{11}\geq\varrho_{33}\geq\varrho_{66}\geq\varrho_{88}\geq\varrho_{55}\geq\varrho_{77},\\
		 &\varrho_{33}\geq\varrho_{11}\geq\varrho_{44}\geq\varrho_{22}\geq\varrho_{77}\geq\varrho_{55}\geq\varrho_{88}\geq\varrho_{66},\,\,\,\,
		 \varrho_{44}\geq\varrho_{22}\geq\varrho_{33}\geq\varrho_{11}\geq\varrho_{88}\geq\varrho_{66}\geq\varrho_{77}\geq\varrho_{55},\\
		 &\varrho_{55}\geq\varrho_{77}\geq\varrho_{66}\geq\varrho_{88}\geq\varrho_{11}\geq\varrho_{33}\geq\varrho_{22}\geq\varrho_{44},\,\,\,\,
		 \varrho_{66}\geq\varrho_{88}\geq\varrho_{55}\geq\varrho_{77}\geq\varrho_{22}\geq\varrho_{44}\geq\varrho_{11}\geq\varrho_{33},\\
		 &\varrho_{77}\geq\varrho_{55}\geq\varrho_{88}\geq\varrho_{66}\geq\varrho_{33}\geq\varrho_{11}\geq\varrho_{44}\geq\varrho_{22},\,\,\,\,
		 \varrho_{88}\geq\varrho_{66}\geq\varrho_{77}\geq\varrho_{55}\geq\varrho_{44}\geq\varrho_{22}\geq\varrho_{33}\geq\varrho_{11}.
	\end{split}
\end{equation}

{\bf Optimal ordering B-A-C:}
\begin{equation}
	\begin{split}
		 &\varrho_{11}\geq\varrho_{22}\geq\varrho_{55}\geq\varrho_{66}\geq\varrho_{33}\geq\varrho_{44}\geq\varrho_{77}\geq\varrho_{88},\,\,\,\,
		 \varrho_{22}\geq\varrho_{11}\geq\varrho_{66}\geq\varrho_{55}\geq\varrho_{44}\geq\varrho_{33}\geq\varrho_{88}\geq\varrho_{77},\\
		 &\varrho_{33}\geq\varrho_{44}\geq\varrho_{77}\geq\varrho_{88}\geq\varrho_{11}\geq\varrho_{22}\geq\varrho_{55}\geq\varrho_{66},\,\,\,\,
		 \varrho_{44}\geq\varrho_{33}\geq\varrho_{88}\geq\varrho_{77}\geq\varrho_{22}\geq\varrho_{11}\geq\varrho_{66}\geq\varrho_{55},\\
		 &\varrho_{55}\geq\varrho_{66}\geq\varrho_{11}\geq\varrho_{22}\geq\varrho_{77}\geq\varrho_{88}\geq\varrho_{33}\geq\varrho_{44},\,\,\,\,
		 \varrho_{66}\geq\varrho_{55}\geq\varrho_{22}\geq\varrho_{11}\geq\varrho_{88}\geq\varrho_{77}\geq\varrho_{44}\geq\varrho_{33},\\
		 &\varrho_{77}\geq\varrho_{88}\geq\varrho_{33}\geq\varrho_{44}\geq\varrho_{55}\geq\varrho_{66}\geq\varrho_{11}\geq\varrho_{22},\,\,\,\,
		 \varrho_{88}\geq\varrho_{77}\geq\varrho_{44}\geq\varrho_{33}\geq\varrho_{66}\geq\varrho_{55}\geq\varrho_{22}\geq\varrho_{11}.
	\end{split}
\end{equation}

{\bf Optimal ordering B-C-A:}
\begin{equation}
	\begin{split}
		 &\varrho_{11}\geq\varrho_{33}\geq\varrho_{55}\geq\varrho_{77}\geq\varrho_{22}\geq\varrho_{44}\geq\varrho_{66}\geq\varrho_{88},\,\,\,\,
		 \varrho_{22}\geq\varrho_{44}\geq\varrho_{66}\geq\varrho_{88}\geq\varrho_{11}\geq\varrho_{33}\geq\varrho_{55}\geq\varrho_{77},\\
		 &\varrho_{33}\geq\varrho_{11}\geq\varrho_{77}\geq\varrho_{55}\geq\varrho_{44}\geq\varrho_{22}\geq\varrho_{88}\geq\varrho_{66},\,\,\,\,
		 \varrho_{44}\geq\varrho_{22}\geq\varrho_{88}\geq\varrho_{66}\geq\varrho_{33}\geq\varrho_{11}\geq\varrho_{77}\geq\varrho_{55},\\
		 &\varrho_{55}\geq\varrho_{77}\geq\varrho_{11}\geq\varrho_{33}\geq\varrho_{66}\geq\varrho_{88}\geq\varrho_{22}\geq\varrho_{44},\,\,\,\,
		 \varrho_{66}\geq\varrho_{88}\geq\varrho_{22}\geq\varrho_{44}\geq\varrho_{55}\geq\varrho_{77}\geq\varrho_{11}\geq\varrho_{33},\\
		 &\varrho_{77}\geq\varrho_{55}\geq\varrho_{33}\geq\varrho_{11}\geq\varrho_{88}\geq\varrho_{66}\geq\varrho_{44}\geq\varrho_{22},\,\,\,\,
		 \varrho_{88}\geq\varrho_{66}\geq\varrho_{44}\geq\varrho_{22}\geq\varrho_{77}\geq\varrho_{55}\geq\varrho_{33}\geq\varrho_{11}.
	\end{split}
\end{equation}

{\bf Optimal ordering C-A-B:}
\begin{equation}
	\begin{split}
		 &\varrho_{11}\geq\varrho_{55}\geq\varrho_{22}\geq\varrho_{66}\geq\varrho_{33}\geq\varrho_{77}\geq\varrho_{44}\geq\varrho_{88},\,\,\,\,
		 \varrho_{22}\geq\varrho_{66}\geq\varrho_{11}\geq\varrho_{55}\geq\varrho_{44}\geq\varrho_{88}\geq\varrho_{33}\geq\varrho_{77},\\
		 &\varrho_{33}\geq\varrho_{77}\geq\varrho_{44}\geq\varrho_{88}\geq\varrho_{11}\geq\varrho_{55}\geq\varrho_{22}\geq\varrho_{66},\,\,\,\,
		 \varrho_{44}\geq\varrho_{88}\geq\varrho_{33}\geq\varrho_{77}\geq\varrho_{22}\geq\varrho_{66}\geq\varrho_{11}\geq\varrho_{55},\\
		 &\varrho_{55}\geq\varrho_{11}\geq\varrho_{66}\geq\varrho_{22}\geq\varrho_{77}\geq\varrho_{33}\geq\varrho_{88}\geq\varrho_{44},\,\,\,\,
		 \varrho_{66}\geq\varrho_{22}\geq\varrho_{55}\geq\varrho_{11}\geq\varrho_{88}\geq\varrho_{44}\geq\varrho_{77}\geq\varrho_{33},\\
		 &\varrho_{77}\geq\varrho_{33}\geq\varrho_{88}\geq\varrho_{44}\geq\varrho_{55}\geq\varrho_{11}\geq\varrho_{66}\geq\varrho_{22},\,\,\,\,
		 \varrho_{88}\geq\varrho_{44}\geq\varrho_{77}\geq\varrho_{33}\geq\varrho_{66}\geq\varrho_{22}\geq\varrho_{55}\geq\varrho_{11}.
	\end{split}
\end{equation}

{\bf Optimal ordering C-B-A:}
\begin{equation}
	\begin{split}
		 &\varrho_{11}\geq\varrho_{55}\geq\varrho_{33}\geq\varrho_{77}\geq\varrho_{22}\geq\varrho_{66}\geq\varrho_{44}\geq\varrho_{88},\,\,\,\,
		 \varrho_{22}\geq\varrho_{66}\geq\varrho_{44}\geq\varrho_{88}\geq\varrho_{11}\geq\varrho_{55}\geq\varrho_{33}\geq\varrho_{77},\\
		 &\varrho_{33}\geq\varrho_{77}\geq\varrho_{11}\geq\varrho_{55}\geq\varrho_{44}\geq\varrho_{88}\geq\varrho_{22}\geq\varrho_{66},\,\,\,\,
		 \varrho_{44}\geq\varrho_{88}\geq\varrho_{22}\geq\varrho_{66}\geq\varrho_{33}\geq\varrho_{77}\geq\varrho_{11}\geq\varrho_{55},\\
		 &\varrho_{55}\geq\varrho_{11}\geq\varrho_{77}\geq\varrho_{33}\geq\varrho_{66}\geq\varrho_{22}\geq\varrho_{88}\geq\varrho_{44},\,\,\,\,
		 \varrho_{66}\geq\varrho_{22}\geq\varrho_{88}\geq\varrho_{44}\geq\varrho_{55}\geq\varrho_{11}\geq\varrho_{77}\geq\varrho_{33},\\
		 &\varrho_{77}\geq\varrho_{33}\geq\varrho_{55}\geq\varrho_{11}\geq\varrho_{88}\geq\varrho_{44}\geq\varrho_{66}\geq\varrho_{22},\,\,\,\,
		 \varrho_{88}\geq\varrho_{44}\geq\varrho_{66}\geq\varrho_{22}\geq\varrho_{77}\geq\varrho_{33}\geq\varrho_{55}\geq\varrho_{11}.
	\end{split}
\end{equation}

We now introduce the incoherent operation protocol for three-qubit systems. Given an initial three-qubit state, the protocol proceeds through the following four steps.

{\bf Step 1.} Calculate the state residual battery capacity $\triangle(\varrho,H)$. If $\triangle(\varrho,H)=0$, the protocol ends. Otherwise, go to the next step.

{\bf Step 2.} Calculate the value of $\mathcal{C}(\varrho_A;H_A)+\mathcal{C}(\varrho_B;H_B)+\mathcal{C}(\varrho_C;H_C)$ and record it as $\text{c}_1$. Determine whether the diagonal ordering belongs to (F1$\sim$F6). If the ordering belongs to (F1$\sim$F6), go to next step. Otherwise, we use the unitary matrices $U_{ij}$ $(1\leq i<j\leq8)$ to adjust the diagonal ordering to the optimal one belonging to (F1$\sim$F6)). Then calculate the sum of subsystems' capacities and record it as $\text{c}_2$. Go to the next step.

{\bf Step 3.} At this time, the diagonal ordering of the density matrix is optimal. For convenience, we still record the state as $\varrho$.
\\
(i)\,If the diagonal elements ordering belongs to (F1), we consider unitary evolutions $U_{12}$, $U_{34}$, $U_{56}$ and $U_{78}$. The subsystem battery capacity of the evolved states are calculated separately, with the maximum value recorded as $\text{c}_3$, and then go to next step. This procedure essentially seeks subsystem capacity enhancement by adjusting $\mathcal{C}_A$, $\mathcal{C}_B$ and $\mathcal{C}_C$ while preserving the values of $\text{IC}_A$ and $\text{IC}_B$. The post-evolution values of $\mathcal{C}_A$, $\mathcal{C}_B$ and $\mathcal{C}_C$ is found in Table I.
\\
(ii)\,If the diagonal elements ordering belongs to (F2), we consider unitary evolutions $U_{13}$, $U_{24}$, $U_{57}$ and $U_{68}$. The subsystem capacity of the evolved states are calculated separately, with the maximum value recorded as $\text{c}_3$, and then go to next step.
\\
(iii)\,If the diagonal elements ordering belongs to (F3), we consider unitary evolutions $U_{12}$, $U_{34}$, $U_{56}$ and $U_{78}$. The subsystem capacity of the evolved states are calculated separately, with the maximum value recorded as $\text{c}_3$, and then go to next step.
\\
(iv)\,If the diagonal elements ordering belongs to (F4), we consider unitary evolutions $U_{15}$, $U_{26}$, $U_{37}$ and $U_{48}$. The subsystem capacity of the evolved states are calculated separately, with the maximum value recorded as $\text{c}_3$, and then go to next step.
\\
(v)\,If the diagonal elements ordering belongs to (F5), we consider unitary evolutions $U_{13}$, $U_{24}$, $U_{57}$ and $U_{68}$. The subsystem capacity of the evolved states are calculated separately, with the maximum value recorded as $\text{c}_3$, and then go to next step.
\\
(vi)\,If the diagonal elements ordering belongs to (F6), we consider unitary evolutions $U_{15}$, $U_{26}$, $U_{37}$ and $U_{48}$. The subsystem capacity of the evolved states are calculated separately, with the maximum value recorded as $\text{c}_3$, and then go to next step.
\\

{\bf Step 4.} Select the maximum value among $\text{c}_1$, $\text{c}_2$ and $\text{c}_3$. Trace back the optimization path through this value. If $\max\{\text{c}_1,\text{c}_2,\text{c}_3\}>\text{c}_1$, it means that our protocol achieves subsystem capacity gain.

\section{Appendix G: Proof of Theorem 5}
\setcounter{equation}{0}
\renewcommand{\theequation}{G\arabic{equation}}
Since the Ising model is a special case of the XXZ model with $\alpha=0$, it suffices to prove Theorem 3 for the XXZ model. For $\mathcal{G}=SU(4)$ and $\mathcal{K}=SU(2)\otimes SU(2)$, we first prove that $\mathcal{G}/\mathcal{K}$ is a Riemannian symmetric space. Consider the decomposition of the Lie algebra of $SU(4)$: $g=m\oplus l$, where $m=\text{span}\{\sigma_i^1\sigma_j^2\}$ and $l=\text{span}\{\sigma_i^1, \sigma_j^2\}\,(i,j\in\{x,y,z\})$. It is readily verified that the Lie bracket relations between $m$ and $l$ satisfy:
\begin{equation}
[m,m]\subseteq l,\,\,\,\,[m,l]\subseteq m,\,\,\,\,[l,l]\subseteq l.
\end{equation}
Therefore, $\mathcal{G}/\mathcal{K}$ is a Riemannian symmetric space. The Lie algebra generated by the control Hamiltonians $H_1, H_2, H_3, H_4$ coincides precisely with $l$, i.e., $\langle H_1, H_2, H_3, H_4\rangle_{LA}=l$, while the coupling term $\sigma_z^{\otimes2}+\alpha(\sigma_x^{\otimes2}+\sigma_y^{\otimes2})\in m$ since it is merely a linear combination of the generators of $m$. Furthermore, the adjoint action of subgroup $\mathcal{K}$ is denoted as $Ad_\mathcal{K}$ such that for any $k\in\mathcal{K}$, $Ad_\mathcal{K}(k)(X)=kXk^{-1}$, $X\in g$. Under the adjoint action $Ad_\mathcal{K}$, $\sigma_z^{\otimes2}+\alpha(\sigma_x^{\otimes2}+\sigma_y^{\otimes2})$ can span all bases of $m$, as it contains coupling terms in multiple directions. In other words, for $\beta\geq0$, $Ad_\mathcal{K}(\beta[\sigma_z^{\otimes2}+\alpha(\sigma_x^{\otimes2}+\sigma_y^{\otimes2})])=m$. Thus the XXZ model Hamiltonian given by Eq.\,(\ref{e17}) satisfies all conditions of Theorem 5 in Ref.\,\cite{nkrb}, which implies that the minimum time required to implement a unitary operation under the XXZ model is the smallest value of $\frac{1}{J}\sum_{i=1}^{3}a_i$, $a_i\geq0$, over the following decompositions,
\begin{equation}	 U=k_1\text{exp}\{\frac{i}{2}(a_1\sigma_x^1\sigma_x^2+a_2\sigma_y^1\sigma_y^2
+a_3\sigma_z^1\sigma_z^2)\}k_2,
\end{equation}
where $k_1, k_2\in\mathcal{K}$.

Note that for a given unitary operation $U$, the decomposition specified by (G2) is not unique, making it highly challenging to find the minimum through exhaustive enumeration of all such decompositions. The authors in Ref.\,\cite{blzh} addressed this issue by taking into account the local unitary invariants within $U(4)$. Specifically, two unitaries $U_1, U_2\in U(4)$ are termed locally equivalent if they satisfy $U_1=V_1U_2V_2$ for some $V_1, V_2\in U(2)\otimes U(2)$. Given $U\in U(4)$, the invariants under such equivalence can be expressed as \cite{blzh,jzjv}:
\begin{equation}
\chi_1=\frac{[\text{Tr}(U^{'})]^2}{16\text{det}U},\,\,\,\,\chi_2=\frac{[\text{Tr}(U^{'})]^2-\text{Tr}(U^{'2})}{4\text{det}U},
\end{equation}
where $U^{'}=(O^\dagger UO)^T(O^\dagger UO)$ with
$$O=\frac{1}{\sqrt{2}}\left(
\begin{array}{cccc}
	1 & 0 & 0 & i\\
	0 & i & 1 & 0\\
	0 & i & -1 & 0\\
	1 & 0 & 0 & -i\\
\end{array}
\right ).$$
Based on the decomposition of $U$ in Eq.\,(G2), $\chi_1$ and $\chi_2$ can be further expressed in terms of $a_1, a_2, a_3$ \cite{blzh,jzjv}: $\chi_1=\omega_1+i\omega_2, \chi_2=\omega_3$, where
\begin{equation}
\begin{split}
\omega_1&=\cos^2a_1\cos^2a_2\cos^2a_3-\sin^2a_1\sin^2a_2\sin^2a_3,\\
\omega_2&=\frac{1}{4}\sin2a_1\sin2a_2\sin2a_3,\\
\omega_3&=4\cos^2a_1\cos^2a_2\cos^2a_3-4\sin^2a_1\sin^2a_2\sin^2a_3-\cos2a_1\cos2a_2\cos2a_3.
\end{split}
\end{equation}
By solving for $a_1, a_2, a_3$ via the local invariants $\omega_1, \omega_2, \omega_3$, one obtains the results independent of the Cartan decomposition (G2).

For the incoherent operations $U_{14}$ and $U_{23}$, we have $\chi_1=-1$ and $\chi_2=-3$ according to Eq.\,(G3). This means that $\omega_1=-1$, $\omega_2=0$ and $\omega_3=-3$. From (G4), we obtain that $a_1=a_2=a_3=\frac{\pi}{2}$. Therefore, the minimum time required to implement operation $U_{14}$ or $U_{23}$ is given by $t^*(U_{14})=t^*(U_{23})=\frac{3\pi}{2J}$. For the other incoherent operations involved in our protocol, we verify through calculation that their corresponding local invariants share identical values. That is to say, $\chi_1=0$, $\chi_2=1$. Then we have $\omega_1=\omega_2=0$ and $\omega_3=1$. Substituting $\omega_1$, $\omega_2$ and $\omega_3$ into (G4), and solving inversely for $a_1$, $a_2$ and $a_3$, we obtain $a_1=\frac{\pi}{2}$ and $a_2=a_3=0$. Consequently, the minimum time required to implement the incoherent operations $U_{12}$, $U_{34}$, $U_{13}$ and $U_{24}$ is
\begin{equation*}
t^*(U_{12})=t^*(U_{34})=t^*(U_{13})=t^*(U_{24})=\frac{1}{J}\sum_{i=1}^{3}a_i=\frac{\pi}{2J}.
\end{equation*}
\end{widetext}
\end{document}